# Reachability Games on Extended Vector Addition Systems with States

Tomáš Brázdil[1], Petr Jančar[2], and Antonín Kučera[1]

[1] Faculty of Informatics, Masaryk University, Botanická 68a,
60200 Brno, Czech Republic. {`brazdil,kucera`}`@fi.muni.cz`
[2] Dept. of Computer Science, FEI, Technical University of Ostrava, 17. listopadu 15, 70833
Ostrava, Czech Republic. `Petr.Jancar@vsb.cz`

**Abstract.** We consider two-player turn-based games with zero-reachability and zero-safety objectives generated by extended vector addition systems with states. Although the problem of deciding the winner in such games is undecidable in general, we identify several decidable and even tractable subcases of this problem obtained by restricting the number of counters and/or the sets of target configurations.

## 1 Introduction

Vector addition systems with states (VASS) are an abstract computational model equivalent to Petri nets (see, e.g., [27, 29]) which is well suited for modelling and analysis of distributed concurrent systems. Roughly speaking, a *k-dimensional VASS*, where $k \geq 1$, is an automaton with a finite control and $k$ unbounded counters which can store non-negative integers. Depending on its current control state, a VASS can choose and perform one of the available transitions. A given transition changes the control state and updates the vector of current counter values by adding a fixed vector of integers which *labels* the transition. For simplicity, we assume that transition labels can increase/decrease each counter at most by one. Since the counters cannot become negative, transitions which attempt to decrease a zero counter are disabled. Configurations of a given VASS are written as pairs $p\mathbf{v}$, where $p$ is a control state and $\mathbf{v} \in \mathbb{N}^k$ a vector of counter values.

In this paper, we consider *extended VASS games* which enrich the modelling power of VASS in two orthogonal ways.

(1) Transition labels can contain *symbolic* components (denoted by $\omega$) whose intuitive meaning is "add an arbitrarily large non-negative integer to a given counter". For example, a single transition $p \to q$ labeled by $(1, \omega)$ represents an infinite number of "ordinary" transitions labeled by $(1, 0), (1, 1), (1, 2), \ldots$ A natural source of motivation for introducing symbolic labels are systems with multiple resources that can be consumed and produced simultaneously by performing a transition. The $\omega$ components can then be conveniently used to model "resource reloading" (see also the example below).

(2) To model the interaction between a system and its environment, the set of control states is split into two disjoint subsets of *controllable* and *environmental* states. Transitions from the controllable and environmental states then correspond to the events generated by the system and its environment, respectively.

Hence, the semantics of a given extended VASS game $\mathcal{M}$ is a possibly infinitely-branching turn-based game $G_\mathcal{M}$ with infinitely many vertices which correspond to the configurations of $\mathcal{M}$. The game $G_\mathcal{M}$ is initiated by putting a token on some configuration $p\mathbf{v}$. The token is then moved from vertex to vertex by two players, □ and ◇, who select transitions in the controllable and environmental configurations according to some strategies. Thus, they produce an infinite sequence of configurations called a *play*. Desired properties of $\mathcal{M}$ can be formalized as *objectives*, i.e., admissible plays. The central problem is the question whether player □ (the system) has a *winning* strategy which ensures that the objective is satisfied for every strategy of player ◇ (the environment). We refer to, e.g., [32, 13, 35] for more comprehensive expositions of results related to games in formal verification. In this paper, we are mainly interested in *zero-safety* objectives (or, dually, *zero-reachability* objectives), consisting of plays where no counter is decreased to zero, i.e., a given system never reaches a situation when some of its resources are insufficient.

As a simple example, consider a workshop which "consumes" wooden sticks, screws, wires, etc., and produces puppets of various kinds which are then sold at the door. From time to time, the manager may decide to issue an order for screws or other supplies, and thus increase their number by a finite but essentially unbounded amount (the manager certainly aims at choosing the "right" number of screws which are needed to produce all puppets that can be sold in next few days). Controllable states can be used to model the actions taken by workshop employees, and environmental states model the behaviour of unpredictable customers. We wonder whether the workshop manager has a strategy which ensures that at least one puppet of each kind is always available for sell, regardless what the unpredictable customers do (the model can of course reflect only selected aspects of customers' behaviour). Note that a winning strategy for the manager must also resolve the symbolic $\omega$ value used to model the order of screws by specifying a *concrete number* of screws that should be ordered.

Technically, we consider extended VASS games with *non-selective* and *selective* zero-reachability objectives, where the set of target configurations that should be reached by player ◇ and avoided by player □ is either $Z$ and $Z_C$, respectively. Here,

– the set $Z$ consists of all $p\mathbf{v}$ such that $\mathbf{v}_\ell = 0$ for some $\ell$ (i.e., some counter is zero);
– the set $Z_C$, where $C$ is a subset of control states, consists of all $p\mathbf{v} \in Z$ such that $p \in C$.

**Our main results** can be summarized as follows:

(a) The problem of deciding the winner in $k$-dimensional extended VASS games (where $k \geq 2$) with $Z$-reachability objectives is in **(k-1)-EXPTIME**.
(b) A finite description of the winning region for each player (i.e., the set of all vertices where the player wins) is computable in $(k-1)$-exponential time.
(c) Winning strategies for both players admit a finite and effectively computable description.

We note that the classical result by Lipton [24] easily implies **EXPSPACE**-hardness (even in the case when player ◇ has no influence). These (decidability) results are complemented by noting the following straightforward undecidability:

(d) The problem of deciding the winner in 2-dimensional VASS games with "ordinary" (non-symbolic) transitions and $Z_C$-reachability objectives is undecidable. The same problem for 3-dimensional *extended* VASS games is *highly* undecidable (beyond the arithmetical hierarchy).

Further, we consider the special case of one-dimensional extended VASS games, where we provide the following (tight) complexity results:

(e) The problem of deciding the winner in one-dimensional extended VASS games with Z-reachability objectives is in **P**. Both players have "counterless" winning strategies constructible in polynomial time.

(f) The problem of deciding the winner in one-dimensional extended VASS games with $Z_C$-reachability objectives is **PSPACE** complete. A finite description of the winning regions is computable in exponential time.

To the best of our knowledge, these are the first positive decidability/tractability results about a natural class of *infinitely branching* turn-based games, and some of the underlying observations are perhaps of broader interest (in particular, we obtain slight generalizations of the "classical" results about self-covering paths achieved by Rackoff [28] and elaborated by Rosier&Yen [30]).

To build a preliminary intuition behind the technical proofs of (a)–(f) presented in Section 3, we give a brief outline of these proofs and sketch some of the crucial insights.

**A proof outline for (a)–(c).** Observe that if the set of environmental states that are controlled by player ◇ is empty, then the existence of a winning strategy for player □ in $p\boldsymbol{v}$ is equivalent to the existence of a *self-covering zero-avoiding path* of the form $p\boldsymbol{v} \to^* q\boldsymbol{u} \to^+ q\boldsymbol{u}'$, where $\boldsymbol{u} \leq \boldsymbol{u}'$ and the counters stay positive along the path. The existence and the size of such paths has been studied in [28, 30] (actually, they mainly consider the existence of an *increasing self-covering path* where $\boldsymbol{u}'$ is *strictly* larger than $\boldsymbol{u}$ in at least one component, and the counters *can* be decreased to zero in the intermediate configurations). One can easily generalize this observation to the case when the set of environmental states is non-empty and show that the existence of a winning strategy for player □ in $p\boldsymbol{v}$ is equivalent to the existence of a *self-covering zero-avoiding tree* initiated in $p\boldsymbol{v}$, which is a finite tree, rooted in $p\boldsymbol{v}$, describing a strategy for player □ where each maximal path (i.e., each branch) is self-covering and zero-avoiding (if player □ follows this strategy, a self-covering zero-avoiding path is necessarily produced after a finite number of steps no matter what player ◇ does).

We show that the existence of a self-covering zero-avoiding tree initiated in a given configuration of a given extended VASS is decidable, and we give some complexity bounds. Let us note that this result is more subtle than it might seem; one can easily show that the existence of a self-covering (but not necessarily zero-avoiding) tree for a given configuration is already *undecidable* (see Appendix A.1 for details).

Our algorithm constructs all *minimal* $p\boldsymbol{v}$ (w.r.t. component-wise ordering) where player □ has a winning strategy. Since this set is necessarily finite, and the winning region of player □ is obviously upwards-closed, we obtain a finite description of the winning region for player □. The algorithm can be viewed as a concrete (but not obvious) instance of a general approach, which is dealt with, e.g., in [33, 10, 11]. First, we compute all control states $p$ such that player □ can win in *some* configuration $p\boldsymbol{v}$. Here, a crucial step is to observe that if this is *not* the case, i.e., player ◇ can win in every $p\boldsymbol{v}$, then player ◇ has a *counterless* winning strategy which depends only on the current control state (since there are only finitely many counterless strategies, they can be tried out one by one). This computation also gives an initial bound $B$ such that for every control state $p$ we have that if player □ wins in *some* $p\boldsymbol{v}$, then he wins in all $p\boldsymbol{v}'$ where $v'_\ell \geq B$ for all indexes (counters) $\ell \in \{1, 2, \ldots, k\}$. Then the algorithm proceeds inductively, explores the situations where at least one counter is less than $B$, computes (bigger) general bounds for the other $k-1$ counters, etc.

A finite description of a strategy for player □ which is winning in every configuration of his winning region is obtained by specifying the moves in all minimal winning configurations (observe that in a non-minimal winning configuration $p(\mathbf{v}+\mathbf{u})$ such that $p\mathbf{v}$ is minimal, player □ can safely make a move $p(\mathbf{v}+\mathbf{u}) \to q(\mathbf{v}'+\mathbf{u})$ where $p\mathbf{v} \to q\mathbf{v}'$ is the move associated to $p\mathbf{v}$). Note that this also resolves the issue with $\omega$ components in transitions performed by player □. Since the number of minimal winning configurations is finite, there is a finite and effectively computable constant $c$ such that player □ never needs to increase a counter by more than $c$ when performing a transition whose label contains a symbolic component (and we can even give a simple "recipe" which gives an optimal choice for the $\omega$ values for every configuration separately).

The winning region of player ◇ is just the complement of the winning region of player □. Computing a finite description of a winning strategy for player ◇ is somewhat trickier and relies on some observations made in the "inductive step" discussed above (note that for player ◇ it is not sufficient to stay in his winning region; he also needs to make some progress in approaching zero in some counter).

**A proof outline for (d).** The undecidability result for 2-dimensional VASS games is obtained by a straightforward reduction of the halting problem for Minsky machines with two counters initialized to zero, which is undecidable [26]. Let us note that this construction is essentially the same as the one for monotonic games presented in [1], and it is included mainly for the sake of completeness. After some minor modifications, the same construction can be also used to establish the undecidability of other natural problems for VASS and extended VASS games, such as boundedness or coverability. The high undecidability result for 3-dimensional extended VASS games is proven by reducing the problem whether a given nondeterministic Minsky machine with two counters initialized to zero has an infinite computation such that the initial instruction is executed infinitely often (this problem is known to be $\Sigma_1^1$-complete [15]). This reduction is also straightforward, but at least it demonstrates that symbolic transitions do bring some extra power (note that for "ordinary" VASS games, a winning strategy for player ◇ in a given $p\mathbf{v}$ can be written as a finite tree, and hence the existence of such a strategy is obviously semidecidable).

**A proof outline for (e)–(f).** The case of one-dimensional extended VASS games with zero-reachability objectives is, of course, simpler than the general case, but our results still require some effort. In the case of Z-reachability objectives, we show that the winning region of player ◇ can be computed as the least fixed point of a monotonic function over a finite lattice. Although the lattice has exponentially many elements, we show that the function reaches the least fixed point only after a quadratic number of iterations. The existence and efficient constructibility of counterless winning strategies is immediate for player □, and we show that the same is achievable for player ◇. The results about $Z_C$-reachability objectives are obtained by applying known results about the emptiness problem for alternating finite automata with one letter alphabet [16] (see also [21]) and the emptiness problem for alternating two-way parity word automata [31], together with some additional observations.

**Related work.** As already mentioned, some of our results and proof techniques use (and generalize) the techniques from [28, 30]. VASS games can be also seen as a special case of *monotonic* games considered in [1], where it is shown that the problem of deciding the winner in monotonic games with reachability objectives is undecidable (see the proof outline for (d) above). Let us note that the results presented in [1] mainly concern the so-called *downward-closed* games, which is a model different from ours. Let us also mention that (extended) VASS games are different from another recently studied model of *branching vector addition systems* [34, 6] which has different semantics and different algorithmic properties (for example, the coverability and boundedness problems for branching vector addition systems are complete for **2-EXPTIME** [6]). We have also mentioned that there are studies of generic procedures applicable to sets of states which are upward-closed w.r.t. a suitable ordering (e.g., [3, 12, 33, 10, 11]); some insight has been needed to show that our setting could be seen as a concrete instance, and further insight has also brought some "algorithmic consequences".

Note that one-dimensional VASS games are essentially one-counter automata where the counter cannot be tested for zero explicitly (that is, there are no transitions enabled only when the counter reaches zero). Such one-counter automata are also called *one-counter nets* because they correspond to Petri nets with just one unbounded place. The models of one-counter automata and one-counter nets have been intensively studied [18, 20, 22, 2, 7, 9, 19, 31, 14]. Many problems about equivalence-checking and model-checking one-counter automata are known to be decidable, but only a few of them are solvable efficiently. From this point of view, we find the polynomial-time result about one-dimensional extended VASS games with $Z$-reachability objectives encouraging.

## 2 Definitions

In this paper, the sets of all integers, positive integers, and non-negative integers are denoted by $\mathbb{Z}$, $\mathbb{N}^{>0}$, and $\mathbb{N}$, respectively. For every finite or countably infinite set $M$, the symbol $M^*$ denotes the set of all finite words (i.e., finite sequences) over $M$. The length of a given word $w$ is denoted by $|w|$ or $length(w)$, and the individual letters in $w$ are denoted by $w(0), w(1), \ldots, w(|w|-1)$. The empty word is denoted by $\varepsilon$, where $|\varepsilon| = 0$. We also use $M^+$ to denote the set $M^* \smallsetminus \{\varepsilon\}$. A *path* in $\mathcal{M} = (M, \rightarrow)$, for a binary relation $\rightarrow \subseteq M \times M$, is a finite or infinite sequence $w = m_0, m_1, \ldots$ such that $m_i \rightarrow m_{i+1}$ for every $i$; we put $length(w) = \omega$ if $w$ is infinite. As above, $w(i)$ denotes the element $m_i$ of $w$; by $w_i$ we denote the (finite or infinite) path $m_i, m_{i+1}, \ldots$. (By writing $w(i) = m$ or $w_i$ we implicitly assume that $length(w) \geq i+1$.) A given $n \in M$ is *reachable* from a given $m \in M$, written $m \rightarrow^* n$, if there is a finite path from $m$ to $n$. A *run* is a maximal path (infinite, or finite which cannot be prolonged). The sets of all finite paths and all runs in $\mathcal{M}$ are denoted by $FPath(\mathcal{M})$ and $Run(\mathcal{M})$, respectively. Similarly, the sets of all finite paths and runs that start in a given $m \in M$ are denoted by $FPath(\mathcal{M}, m)$ and $Run(\mathcal{M}, m)$, respectively.

**Definition 1 (Game).** *A* game *is a tuple $G = (V, \mapsto, (V_\square, V_\diamond))$ where $V$ is a finite or countably infinite set of* vertices, *$\mapsto \subseteq V \times V$ is an* edge relation, *and $(V_\square, V_\diamond)$ is a partition of $V$.*

A game is played by two players, $\square$ and $\diamond$, who select the moves in the vertices of $V_\square$ and $V_\diamond$, respectively. Let $\odot \in \{\square, \diamond\}$. A *strategy* for player $\odot$ is a (partial) function which to each $wv \in V^*V_\odot$ assigns a vertex $v'$ such that $v \mapsto v'$ if there is any. The set of all strategies for player $\square$ and player $\diamond$ is denoted by $\Sigma$ and $\Pi$, respectively. We say that a strategy $\tau$ is *memoryless* if $\tau(wv)$ depends just on the last vertex $v$. In the rest of this paper, we consider memoryless strategies as (partial) functions from $V_\odot$ to $V$.

A *winning objective* is a set of runs $\mathcal{W} \subseteq Run(G)$. Every pair of strategies $(\sigma, \pi) \in \Sigma \times \Pi$ and every initial vertex $v \in V$ determine a unique run $G^{(\sigma,\pi)}(v) \in Run(G, v)$ which is called a *play*. We say that a strategy $\sigma \in \Sigma$ is $\mathcal{W}$-*winning* (for player $\square$) in a given $v \in V$ if for every $\pi \in \Pi$ we have that $G^{(\sigma,\pi)}(v) \in \mathcal{W}$. Similarly, a strategy $\pi \in \Pi$ is $\mathcal{W}$-winning for player $\diamond$ if for every $\sigma \in \Sigma$ we have that $G^{(\sigma,\pi)}(v) \in \mathcal{W}$. The set of all vertices where player $\odot$ has a $\mathcal{W}$-winning strategy is called the *winning region* of player $\odot$ and denoted by $Win(\odot, \mathcal{W})$.

In this paper, we only consider *reachability* and *safety* objectives, which are specified by a subset of target vertices that should or should not be reached by a run, respectively. Formally, for a given $T \subseteq V$ we define the sets of runs $\mathcal{R}(T)$ and $\mathcal{S}(T)$, where

- $\mathcal{R}(T) = \{w \in Run(G) \mid w(i) \in T \text{ for some } i\}$,
- $\mathcal{S}(T) = \{w \in Run(G) \mid w(i) \notin T \text{ for all } i\}$.

We note that $\mathcal{R}(T) = Run(G) \smallsetminus \mathcal{S}(T)$, and the games with reachability and safety objectives are *determined*, i.e., $Win(\square, \mathcal{S}(T)) = V \smallsetminus Win(\diamond, \mathcal{R}(T))$; moreover, each player has a *memoryless* winning strategy in every vertex of his winning region[3].

**Definition 2 (extended VASS game).** *Let $k \in \mathbb{N}^{>0}$. A $k$-dimensional* vector addition system with states (VASS) *is a tuple $\mathcal{M} = (Q, T, \alpha, \beta, \delta)$ where $Q \neq \emptyset$ is a finite set of* control states, $T \neq \emptyset$ *is a finite set of* transitions, $\alpha : T \to Q$ *and* $\beta : T \to Q$ *are the* source *and* target *mappings, and* $\delta : T \to \{-1, 0, 1\}^k$ *is a transition displacement* labeling. *For technical convenience, we assume that for every $q \in Q$ there is some $t \in T$ such that $\alpha(t) = q$.*

*An* extended VASS (eVASS for short) *is a VASS where the transition displacement labeling is a function $\delta : T \to \{-1, 0, 1, \omega\}^k$.*

*A VASS game (or eVASS game) is a tuple $\mathcal{M} = (Q, (Q_\square, Q_\diamond), T, \alpha, \beta, \delta)$ where $(Q, T, \alpha, \beta, \delta)$ is a VASS (or eVASS) and $(Q_\square, Q_\diamond)$ is a partition of $Q$.*

A *configuration* of $\mathcal{M}$ is an element of $Q \times \mathbb{N}^k$. We write $p\boldsymbol{v}$ instead of $(p, \boldsymbol{v})$, and the $\ell$-th component of $\boldsymbol{v}$ is denoted by $\boldsymbol{v}_\ell$. For a given transition $t \in T$, we write $t : p \to q$ to indicate that $\alpha(t) = p$ and $\beta(t) = q$, and $p \xrightarrow{\boldsymbol{v}} q$ to indicate that $p \to q$ and $\delta(t) = \boldsymbol{v}$. A transition $t \in T$ is *enabled* in a configuration $p\boldsymbol{v}$ if $\alpha(t) = p$ and for every $1 \leq \ell \leq k$ such that $\delta(t)_\ell = -1$ we have $\boldsymbol{v}_\ell \geq 1$.

Every $k$-dimensional eVASS game $\mathcal{M} = (Q, (Q_\square, Q_\diamond), T, \alpha, \beta, \delta)$ induces a unique infinite-state game $G_\mathcal{M}$ where $Q \times \mathbb{N}^k$ is the set of vertices partitioned into $Q_\square \times \mathbb{N}^k$ and $Q_\diamond \times \mathbb{N}^k$, and $p\boldsymbol{v} \mapsto q\boldsymbol{u}$ iff the following condition holds:

- there is a transition $t \in T$ enabled in $p\boldsymbol{v}$ such that $\beta(t) = q$ and for every $1 \leq \ell \leq k$ we have that $\boldsymbol{u}_\ell - \boldsymbol{v}_\ell$ is either non-negative or equal to $\delta(t)_\ell$, depending on whether $\delta(t)_\ell = \omega$ or not, respectively.

Note that any play can get stuck only when a counter is zero, because there is at least one enabled transition otherwise.

In this paper, we are interested in VASS and eVASS games with *non-selective* and *selective* zero-reachability objectives. Formally, for every $C \subseteq Q$ we define the set

$$Z_C = \{p\boldsymbol{v} \in Q \times \mathbb{N}^k \mid p \in C \text{ and } \boldsymbol{v}_i = 0 \text{ for some } 0 \leq i \leq k\}$$

and we also put $Z = Z_Q$. Selective (or non-selective) zero-reachability objectives are reachability objectives where the set $T$ of target configurations is equal to $Z_C$ for some $C \subseteq Q$ (or to $Z$, respectively).

---
[3] In this paper, we consider infinitely-branching games with countable state space. The determinacy result of Martin [25] holds also for this type of games, and memoryless determinacy can be easily established by standard methods.

As we have already noted, our games with reachability objectives are memoryless determined and this result of course applies also to eVASS games with zero-reachability objectives. However, since eVASS games have infinitely many vertices, not all memoryless strategies are finitely representable. In this paper we will often deal with a simple form of memoryless strategies, where the decision is independent of the current counter values; such strategies are called *counterless strategies*.

**Definition 3.** *Given (the game induced by) an eVASS* $M = (Q, (Q_\square, Q_\diamond), T, \alpha, \beta, \delta)$, *a strategy $\tau$ of player* $\odot \in \{\square, \diamond\}$ *is* counterless *if it determines a (fixed) transition $t_p$ for each $p \in Q_\odot$, together with (fixed) values $c_\ell \in \mathbb{N}$ for all those $\ell$ for which $\delta(t_p)_\ell = \omega$, so that $\tau(pv)$ is the configuration arising by performing $t_p$ where $\omega$'s are instantiated with $c_\ell$.*

## 3 VASS and eVASS games with zero-reachability objectives

In this section, we analyze VASS and eVASS games with zero-reachability objectives. We first note the problems of our interest are undecidable for $\mathcal{R}(Z_C)$ objectives; this can be shown by (simple modifications of) standard techniques.

**Proposition 4.** *The problem of deciding the winner in 2-dimensional VASS games with $\mathcal{R}(Z_C)$ objectives is undecidable. For 3-dimensional eVASS games, the same problem is highly undecidable (i.e., beyond the arithmetical hierarchy).*

Let us note that Proposition 4 cannot be extended to one-dimensional eVASS games, which are analyzed later in Section 3.1. Further, by some trivial modifications of the proof of Proposition 4 we also get the undecidability of the boundedness/coverability problems for 2-dimensional VASS games (a given configuration $pv$ is *bounded* if player $\diamond$ has a strategy such that all counters stay bounded for every strategy of player $\square$; similarly, a configuration $qv$ is *coverable* from an initial configuration $pv$ if player $\diamond$ has a strategy such that a configuration of the form $qv'$, where $v' \geq v$, is reached for every strategy of player $\square$). The details are given in Appendix A.1.

Now we turn our attention to $\mathcal{R}(Z)$ objectives. For the rest of this section, we fix a $k$-dimensional eVASS game $M = (Q, (Q_\square, Q_\diamond), T, \alpha, \beta, \delta)$. Since we are interested only in $\mathcal{R}(Z)$ objectives, we may safely assume that every transition $p \xrightarrow{v} q$ of $M$ where $p \in Q_\diamond$ satisfies $v_\ell \neq \omega$ for every $1 \leq \ell \leq k$ (if there are some $\omega$-components in $v$, they can be safely replaced with 0). We also use $d$ to denote the *branching degree* of $M$, i.e, the least number such that every $q \in Q$ has at most $d$ outgoing transitions.

We also use the partial order $\leq$ on the set of configurations of $M$ defined by $pu \leq qv$ iff $p = q$ and $u \leq v$ (componentwise). For short, we write $Win_\diamond$ instead of $Win(\diamond, \mathcal{R}(Z))$ and $Win_\square$ instead of $Win(\square, \mathcal{S}(Z))$. Obviously, if player $\diamond$ has a winning strategy in $qv$, then he can use "essentially the same" strategy in $qu$ for every $u \leq v$ (behaving in $q'v'$ as previously in $q'(v' + v - u)$, which results in reaching 0 in some counter possibly even earlier). Similarly, if $qv \in Win_\square$ then $qu \in Win_\square$ for every $u \geq v$. Thus, we obtain the following:

**Proposition 5.** $Win_\diamond$ *is downwards closed and* $Win_\square$ *is upwards closed w.r.t.* $\leq$.

A direct corollary to Proposition 5 is that the set $Win_\square$ is finitely representable by its subset $Min_\square$ of *minimal elements* (note that $Min_\square$ is necessarily finite because there is no infinite subset of $\mathbb{N}^k$ with pairwise incomparable elements, as Dickson's Lemma shows). Technically, it is convenient to consider also *symbolic configurations* of $M$ which are introduced in the next definition.

**Definition 6.** *A* symbolic configuration *is a pair $q\boldsymbol{v}$ where $q \in Q$ and $\boldsymbol{v} \in (\mathbb{N} \cup \{\omega\})^k$. We say that a given index $\ell \in \{1, 2, \ldots, k\}$ is* precise *in $q\boldsymbol{v}$ if $\boldsymbol{v}_\ell \in \mathbb{N}$, otherwise it is* symbolic *in $q\boldsymbol{v}$. The* precision *of $q\boldsymbol{v}$, denoted by $P(q\boldsymbol{v})$, is the number of indexes that are precise in $q\boldsymbol{v}$. We say that a configuration $p\boldsymbol{u}$* matches *a symbolic configuration $q\boldsymbol{v}$ if $p = q$ and $\boldsymbol{u}_\ell = \boldsymbol{v}_\ell$ for every $\ell$ precise in $q\boldsymbol{v}$. Similarly, we say that $p\boldsymbol{u}$ matches $q\boldsymbol{v}$* above *a given bound $B \in \mathbb{N}$ if $p\boldsymbol{u}$ matches $q\boldsymbol{v}$ and $\boldsymbol{u}_\ell \geq B$ for every $\ell$ symbolic in $q\boldsymbol{v}$.*

We extend the set $Win_\square$ by all symbolic configurations $q\boldsymbol{v}$ such that some configuration matching $q\boldsymbol{v}$ belongs to $Win_\square$. Similarly, the set $Win_\diamond$ is extended by all symbolic configurations $q\boldsymbol{v}$ such that all configurations matching $q\boldsymbol{v}$ belong to $Win_\diamond$ (note that every symbolic configuration belongs either to $Win_\square$ or to $Win_\diamond$). We also extend the previously fixed ordering on configurations to symbolic configurations by stipulating that $\omega \leq \omega$ and $n < \omega$ for all $n \in \mathbb{N}$. Obviously, this extension does not influence the set $Min_\square$, and the winning region $Win_\diamond$ can be now represented by its subset $Max_\diamond$ of all maximal elements, which is necessarily finite.

Our ultimate goal is to compute the sets $Min_\square$ and $Max_\diamond$. Since our reachability games are determined, it actually suffices to compute just one of these sets. In the following we show how to compute $Min_\square$.

We start with an important observation about winning strategies for player $\square$, which in fact extends the "classical" observation about self-covering paths in vector addition systems presented in [28]. Let $q \in Q$ be such that $q\boldsymbol{v} \in Win_\square$ for some $\boldsymbol{v}$, i.e., $q(\omega, \ldots, \omega) \in Win_\square$. This means that there is a strategy of player $\square$ that prevents unbounded decreasing of the counters; we find useful to represent the strategy by a finite *unrestricted self-covering tree for $q$*. The word "unrestricted" reflects the fact that we also consider configurations with negative and symbolic counter values. More precisely, an unrestricted self-covering tree for $q$ is a finite tree $\mathcal{T}$ whose nodes are labeled by the elements of $Q \times (\mathbb{Z} \cup \{\omega\})^k$ satisfying the following ($\omega$ is treated in the standard way, i.e., $\omega + \omega = \omega + c = \omega$ for every $c \in \mathbb{Z}$).

- The root of $\mathcal{T}$ is labeled by $q(0, \ldots, 0)$.
- If $n$ is a non-leaf node of $\mathcal{T}$ labeled by $p\boldsymbol{u}$, then
    - if $p \in Q_\square$, then $n$ has only one successor labeled by some $r\boldsymbol{t}$ such that $\mathcal{M}$ has a transition $p \xrightarrow{\boldsymbol{v}} r$ where $\boldsymbol{t} = \boldsymbol{u} + \boldsymbol{v}$;
    - if $p \in Q_\diamond$, then there is a one-to-one correspondence between the successors of $n$ and transitions of $\mathcal{M}$ of the form $p \xrightarrow{\boldsymbol{v}} r$. The node which corresponds to a transition $p \xrightarrow{\boldsymbol{v}} r$ is labeled by $r\boldsymbol{t}$ where $\boldsymbol{t} = \boldsymbol{u} + \boldsymbol{v}$.
- If $n$ is a leaf of $\mathcal{T}$ labeled by $p\boldsymbol{u}$, then there is another node $m$ (where $m \neq n$) on the path from the root of $\mathcal{T}$ to $n$ which is labeled by $p\boldsymbol{t}$ for some $\boldsymbol{t} \leq \boldsymbol{u}$.

The next lemma bounds the depth of such a tree.

**Lemma 7.** *Let $q(\omega, \ldots, \omega) \in Win_\square$ (i.e., $q\boldsymbol{v} \in Win_\square$ for some $\boldsymbol{v}$). Then there is an unrestricted self-covering tree for $q$ of depth at most $f(|Q|, d, k) = 2^{(d-1) \cdot |Q|} \cdot |Q|^{c \cdot k^2}$, where $c$ is a fixed constant independent of $\mathcal{M}$ (and $d$ is the branching degree of $\mathcal{M}$).*

Lemma 7 thus implies that if $q(\omega, \ldots, \omega) \in Win_\square$, then $q\boldsymbol{u} \in Win_\square$ for all $\boldsymbol{u}$ with $\boldsymbol{u}_\ell \geq f(|Q|, d, k)$ for all $\ell \in \{1, 2, \ldots, k\}$ (recall that each counter can be decreased at most by one in a single transition). The next lemma shows that we can compute the set of all $q \in Q$ such that $q(\omega, \ldots, \omega) \in Win_\square$ (the lemma is formulated "dually", i.e., for player $\diamond$).

**Lemma 8.** *The set of all $q \in Q$ such that $q(\omega, \ldots, \omega) \in \text{Win}_\Diamond$ is computable in space bounded by a polynomial function $g(|Q|, d, k)$.*

An important observation, which is crucial in our proof of Lemma 8 and perhaps interesting on its own, is that if $q(\omega, \ldots, \omega) \in \text{Win}_\Diamond$, then player $\Diamond$ has a *counterless* strategy which is winning in every configuration matching $q(\omega, \ldots, \omega)$. The details are given in Appendix A.2.

To sum up, we can compute the set of all $q(\omega, \ldots, \omega) \in \text{Win}_\Box$ and a bound $B$ which is "safe" for all $q(\omega, \ldots, \omega) \in \text{Win}_\Box$ in the sense that all configurations matching $q(\omega, \ldots, \omega)$ above $B$ belong to $\text{Win}_\Box$. Intuitively, the next step is to find out what happens if one of the counters, say the first one, stays bounded by $B$. Obviously, there is the *least* $j \leq B$ such that $q(j, \omega, \ldots, \omega) \in \text{Win}_\Box$, and there is a bound $D > B$ such that all configurations matching $q(j, \omega, \ldots, \omega)$ above $D$ belong to $\text{Win}_\Box$. If we manage to compute the minimal $j$ (also for the other counters, not just for the first one) and the bound $D$, we can go on and try to bound *two* counters simultaneously by $D$, find the corresponding minima, and construct a new "safe" bound. In this way, we eventually bound all counters and compute the set $\text{Min}_\Box$. In our next definition, we introduce some notions that are needed to formulate the above intuition precisely. (Recall that $P(q\mathbf{v})$ gives the number of precise, i.e. non-$\omega$, elements of $\mathbf{v}$.)

**Definition 9.** *For a given $0 \leq j \leq k$, let $\text{SymMin}_\Box^j$ be the set of all minimal $q\mathbf{v} \in \text{Win}_\Box$ such that $P(q\mathbf{v}) = j$. Further, let $\text{SymMin}_\Box = \bigcup_{i=0}^k \text{SymMin}_\Box^i$. We say that a given $B \in \mathbb{N}$ is safe for precision $j$, where $0 \leq j \leq k$, if for every $q\mathbf{v} \in \bigcup_{i=0}^j \text{SymMin}_\Box^i$ we have that $\mathbf{v}_\ell \leq B$ for every precise index $\ell$ in $\mathbf{v}$, and every configuration matching $q\mathbf{v}$ above $B$ belongs to $\text{Win}_\Box$.*

Obviously, every $\text{SymMin}_\Box^j$ (and hence also $\text{SymMin}_\Box$) is finite, and $\text{Min}_\Box = \text{SymMin}_\Box^k$. Also observe that $\text{SymMin}_\Box^0$ is computable in time exponential in $|Q|$ and $k$ by Lemma 8, and a bound which is safe for precision 0 is computable in polynomial time by Lemma 7. Now we design an algorithm which computes $\text{SymMin}_\Box^{j+1}$ and a bound safe for precision $j+1$, assuming that $\text{SymMin}_\Box^i$ for all $i \leq j$ and a bound safe for precision $j$ have already been computed. A detailed description of the algorithm and the associated proofs can be found in Appendix A.3.

*Remark 10.* To prevent possible confusions, let us note explicitly that the set $\text{SymMin}_\Box^{j+1}$ cannot be obtained from $\text{SymMin}_\Box^j$ simply by considering all $q\mathbf{v} \in \text{SymMin}_\Box^j$ and replacing some $\mathbf{v}_\ell$, where $\ell$ is symbolic in $q\mathbf{v}$, with some concrete value. The set $\text{SymMin}_\Box^{j+1}$ can be substantially richer. For example, if $\text{SymMin}_\Box^j$ contains $p(1, \omega)$ and $p(\omega, 1)$, then $\text{SymMin}_\Box^{j+1}$ surely contains some elements obtained by replacing the $\omega$'s with some concrete values, say $p(1, 10)$ and $p(12, 1)$, but it can contain also other incomparable elements such as $p(2, 9), p(3, 8), \ldots$

**Lemma 11.** *Let $0 \leq j < k$, and let us assume that $\bigcup_{i=0}^j \text{SymMin}_\Box^i$ has already been computed, together with some bound $B \in \mathbb{N}$ which is safe for precision $j$. Then $\text{SymMin}_\Box^{j+1}$ is computable in time exponential in $|Q| \cdot B^{j+1}$, $d$, and $k-j-1$, and the bound $B + f(|Q| \cdot B^{j+1}, d, k-j-1)$ is safe for precision $j + 1$ (here $f$ is the function of Lemma 7 and $d$ is the branching degree of $\mathcal{M}$).*

Now we can easily evaluate the total complexity of computing $\text{SymMin}_\Box$ (and hence also $\text{Min}_\Box$). If we just examine the recurrence of Lemma 11, we obtain that the set $\text{SymMin}_\Box$ is computable in $k$-exponential time. However, we can actually decrease the height of the tower of exponentials by one when we incorporate the results presented in Section 3.1, which imply that for one-dimensional eVASS games, the depth of an unrestricted self-covering tree can be bounded by a *polynomial* in $|Q|$ and $d$, and the set of all $q \in Q$ such that $q(\omega) \in Win_\diamond$ is computable in *polynomial time*. Hence, we actually need to "nest" Lemma 11 only $k-1$ times. Thus, relying on the results of Section 3.1, we obtain the following (where 0-exponential time denotes polynomial time):

**Theorem 12.** *(Given a k-dimensional eVASS), the set $Min_\square$ is computable in $(k-1)$-exponential time.*

Let us note a substantial improvement in complexity would be achieved by improving the bound presented in Lemma 7. Actually, it is not so important what is the depth of an unrestricted self-covering tree, but what are the minimal numbers that allow for applying the strategy described by this tree without reaching zero (i.e., what is the maximal decrease of a counter in the tree). A more detailed complexity analysis based on the introduced parameters reveals that if the maximal counter decrease was just polynomial in the number of control states (which is our conjecture), the complexity bound of Theorem 12 would be *polynomial* for every fixed dimension $k$ (see also Section 4).

Note that after computing the set $Min_\square$, we can easily compute a finite description of a strategy $\sigma$ for player $\square$ which is winning in every configuration of $Win_\square$. For every $pv \in Min_\square$ such that $p \in Q_\square$, we put $\sigma(pv) = qv'$, where $qv'$ is (some) configuration such that $qv' \geq qt$ for some $qt \in Min_\square$. Note that there must be at least one such $qv'$ and it can be computed effectively. For every configuration $pu$ such that $pu \geq pv$ for some $pv \in Min_\square$, we put $\sigma(pu) = q(v'+u-v)$ where $\sigma(pv) = qv'$ (if there are more candidates for $pv$, any of them can be chosen). It is easy to see that $\sigma$ is winning in every configuration of $Win_\square$. Also observe that if we aim at constructing a winning strategy for player $\square$ which minimizes the concrete numbers used to substitute $\omega$'s, we can use $Min_\square$ to construct an "optimal" choice of the values which are sufficient (and necessary) to stay in the winning region of player $\square$.

### 3.1 One-dimensional VASS and eVASS games with zero-reachability objectives.

In this subsection, we present a complete solution for the special case of one-dimensional VASS and eVASS games with zero-reachability objectives.

For the rest of this section, we fix a one-dimensional eVASS game $\mathcal{M} = (Q, (Q_\square, Q_\diamond), T, \alpha, \beta, \delta)$ and $C \subseteq Q$. For every $i \in \mathbb{N}$, let $Win_\diamond(C, i) = \{p \in Q \mid p(i) \in Win(\diamond, \mathcal{R}(Z_C))\}$. It is easy to see that if $Win_\diamond(C, i) = Win_\diamond(C, j)$ for some $i, j \in \mathbb{N}$, then also $Win_\diamond(C, i+1) = Win_\diamond(C, j+1)$. Let $m_C$ be the least $i \in \mathbb{N}$ such that $Win_\diamond(C, i) = Win_\diamond(C, j)$ for some $j > i$, and let $n_C$ be the least $i > 0$ such that $Win_\diamond(C, m_C) = Win_\diamond(C, m_C+i)$. Obviously, $m_c + n_c \leq 2^{|Q|}$ and for every $i \geq m_c$ we have that $Win_\diamond(C, i) = Win_\diamond(C, m_C + ((i-m_C) \mod n_C))$. Hence, the winning regions of both players are fully characterized by all $Win_\diamond(C, i)$, where $0 \leq i < m_C + n_C$.

The selective subcase in analyzed in the following theorem. The **PSPACE** lower bound is obtained by reducing the emptiness problem for alternating finite automata (AFA) with one letter alphabet, which is known to be **PSPACE** complete [16] (see also [21] for a simpler proof). The **PSPACE** upper bound follows by employing the result of [31] which says that the emptiness problem for alternating two-way parity word automata (2PWA) is in **PSPACE** (we would like to thank Olivier Serre for providing us with relevant references). The effective constructability of the winning strategies for player $\square$ and player $\diamond$ follows by applying the results on non-selective termination presented below. The details are given in Appendix A.4.

**Theorem 13.** *The problem whether $p(i) \in \mathit{Win}(\Diamond, \mathcal{R}(Z_C))$ is **PSPACE**-complete. Further, there is a strategy $\sigma$ winning for player $\Box$ in every configuration of $\mathit{Win}(\Box, \mathcal{S}(Z_C))$ such that for all $p \in Q_\Box$ and $i \geq m_C$ we have that $\sigma(p(i)) = \sigma(p(m_C + ((i - m_C) \mod n_C)))$. The numbers $m_C, n_C$ and the tuple of all $\mathit{Win}_\Diamond(C, i)$ and $\sigma(p(i))$, where $0 \leq i < m_C + n_C$ and $p \in Q_\Box$, are constructible in time exponential in $|\mathcal{M}|$.*

In the non-selective subcase, the situation is even better. The winning regions for both players are monotone, which means that $m_Q \leq |Q|$ and $n_Q = 1$. Further, all of the considered problems are solvable in polynomial time.

**Theorem 14.** *The problem whether $p(i) \in \mathit{Win}(\Diamond, \mathcal{R}(Z))$ is in **P**. Further, there are counterless strategies $\sigma$ and $\pi$ such that $\sigma$ is winning for player $\Box$ in every configuration of $\mathit{Win}(\Box, \mathcal{S}(Z))$ and $\pi$ is winning for player $\Diamond$ in every configuration of $\mathit{Win}(\Diamond, \mathcal{R}(Z))$. The tuple of all $\mathit{Win}_\Diamond(Q, i)$, $\sigma(p)$, and $\pi(q)$, where $0 \leq i \leq m_C$, $p \in Q_\Box$, and $q \in Q_\Diamond$, is constructible in time polynomial in $|\mathcal{M}|$.*

## 4 Conclusions, future work

Technically, the most involved result presented in this paper is Theorem 12. This decidability result is not obvious, because most of the problems related to formal verification of Petri nets (equivalence-checking, model-checking, etc.) are undecidable [8, 17, 23, 5]. Since the upper complexity bound given in Theorem 12 is complemented only by the **EXPSPACE** lower bound, which is easily derivable from [24], there is a complexity gap which constitutes an interesting challenge for future work. We conjecture that for a suitable (and reasonable) choice of parameters, one might even obtain *fixed parameter tractability* of the problem. So far, we have not found any arguments against the hypothesis that the problem is tractable, i.e., solvable in polynomial time, even for a *fixed number of counters* (note that the **EXPSPACE** lower bound does not hold for a fixed number of counters).

# A Proofs

In this section we give full proofs of our results together with some auxiliary observations.

## A.1 A proof of Proposition 4

**Proposition 4.** *The problem of deciding the winner in* 2*-dimensional VASS games with* $\mathcal{R}(Z_C)$ *objectives is undecidable. For* 3*-dimensional eVASS games, the same problem is* highly *undecidable (beyond the arithmetical hierarchy).*

*Proof.* The first claim is proven by reducing the halting problem for Minsky machines. A Minsky machine with two counters $c_1, c_2$ is a finite sequence of numbered instructions $1{:}ins_1, \cdots, m{:}ins_m$, where $ins_m = \texttt{halt}$, and for every $1 \leq i < m$ we have that $ins_i$ is either of the form $\texttt{inc } c_j; \texttt{ goto } k$ (type I instructions) or $\texttt{if } c_j{=}0 \texttt{ then goto } k \texttt{ else dec } c_j; \texttt{ goto } n$ (type II instructions). Here $j \in \{1, 2\}$. The problem whether a given Minsky machine with two counters initialized to 0 halts (i.e., executes $\texttt{halt}$ in a finite computation initialized by $ins_1$) is undecidable [26]. For a given Minsky machine $M$ with $m$ instructions, we construct a 2-dimensional VASS game as follows. For every $1 \leq i \leq m$ we add a control state $q_i \in Q_\Diamond$. Further, for every type I instruction $\ell_i : \texttt{inc } c_j; \texttt{ goto } k$ we add a transition $q_i \to q_k$ labeled by $(u_1, u_2)$, where $u_j = 1$ and $u_1 + u_2 = 1$. For every type II instruction $\texttt{if } c_j{=}0 \texttt{ then goto } k \texttt{ else dec } c_j; \texttt{ goto } n$ we add control states $p_i, r_i \in Q_\Box$, and transitions

$$q_i \xrightarrow{(u_1,u_2)} q_j, \quad q_i \xrightarrow{(0,0)} p_i, \quad p_i \xrightarrow{(0,0)} q_k, \quad p_i \xrightarrow{(u_1,u_2)} r_i, \quad r_i \xrightarrow{(0,0)} r_i,$$

where $u_j = -1$ and $u_1 + u_2 = -1$. Finally, we add transitions $q_m \xrightarrow{(-1,0)} q_m$ and $q_m \xrightarrow{(0,-1)} q_m$. Now one can easily check that $M$ halts iff $q_1(0,0) \in Win(\Diamond, \mathcal{R}(Z_{\{q_m\}}))$.

A proof of the second claim is obtained by reducing the problem whether a given *nondeterministic* Minsky machine with two counters initialized to zero has an infinite computation such that the initial instruction is executed infinitely often (this problem is known to be $\Sigma_1^1$-complete [15]). Formally, a nondeterministic Minsky machine with two counters $c_1, c_2$ is a finite sequence of numbered instructions $1{:}ins_1, \cdots, m{:}ins_m$, where each $ins_i$ is of one of the following forms (where $j \in \{1, 2\}$):

- $c_j := c_j{+}1; \texttt{ goto } k$ (type I instructions);
- $\texttt{if } c_j{=}0 \texttt{ then goto } k \texttt{ else } c_j := c_j{-}1; \texttt{goto } n$ (type II instructions);
- $\texttt{goto } \{k \texttt{ or } n\}$ (type III instructions).

Here the indexes $k, n$ range over $\{1, \cdots, m\}$. Note that we may safely assume that the first instruction is of the form $1 : c_1 := c_1{+}1; \texttt{ goto } 2$. For a given nondeterministic Minsky machine $M$ with $m$ instructions, we construct a 3-dimensional eVASS game as follows. For every $1 \leq i \leq m$ we add a control state $q_i \in Q_\Box$. Further, we add a transition $q_1 \xrightarrow{(1,0,\omega)} q_2$, and for every $2 \leq i \leq m$ we add either the transition $q_i \xrightarrow{(u_1,u_2,-1)} q_k$ where $u_j = 1$ and $u_1 + u_2 = 1$, or control states $p_i, r_i \in Q_\Diamond$ together with transitions

$$q_i \xrightarrow{(u_1,u_2,-1)} q_j, \quad q_i \xrightarrow{(0,0,-1)} p_i, \quad p_i \xrightarrow{(0,0,-1)} q_k, \quad p_i \xrightarrow{(u_1,u_2,-1)} r_i, \quad r_i \xrightarrow{(0,0,-1)} r_i,$$

or transitions $q_i \xrightarrow{(0,0,-1)} q_k, q_i \xrightarrow{(0,0,-1)} q_n$, depending on whether $ins_i$ is a type I, type II, or type III instruction, respectively. Note that the third counter can be incremented by an arbitrarily large value whenever the control state $q_1$ is visited. Hence, if $M$ has an infinite computation such that $ins_1$ is executed infinitely often, then player $\square$ can win the game initiated in $q_1(0, 0, 0)$ by simulating this infinite computation and "guessing" the number of steps that are needed to revisit $q_1$. It is also easy to see that if $M$ has no such computation, then player $\diamond$ can win. $\square$

Note that the 2-dimensional VASS game constructed in the proof of the first claim has the property that $M$ halts iff player $\diamond$ has a strategy such that for every strategy of player $\square$ the play initiated in $q_1(0, 0)$ reaches a configuration $q_m\boldsymbol{u}$ where $\boldsymbol{u} \geq (0, 0)$. Hence, the coverability problem for 2-dimensional VASS games is also undecidable. Similarly, if we change the transition $r_i \xrightarrow{(0,0)} r_i$ into $r_i \xrightarrow{(1,1)} r_i$, we obtain that $M$ is space-bounded iff player $\diamond$ has a strategy such that for every strategy of player $\square$ the play initiated in $q_1(0, 0)$ is bounded. This means that the boundedness problem for 2-dimensional VASS games is undecidable. Finally, let us prove the observation mentioned in Section 1, which says that the existence of a self-covering (but not necessarily zero-avoiding) tree in for a given eVASS configuration is undecidable. To prevent possible confusions, let us first clarify what we mean by a self-covering tree for an eVASS configuration.

Let $\mathcal{M} = (Q, (Q_\square, Q_\diamond), T, \alpha, \beta, \delta)$ be a $k$-dimensional eVASS game and $q\boldsymbol{v}$ a configuration of $\mathcal{M}$. A *self-covering tree* for $q\boldsymbol{v}$ is a finite tree $T$ whose nodes are labeled by the elements of $Q \times \mathbb{N}^k$ satisfying the following:

- The root of $T$ is labeled by $q\boldsymbol{v}$.
- If $n$ is an inner node of $T$ labeled by $p\boldsymbol{u}$, then
  - if $p \in Q_\square$, then $n$ has exactly one successor labeled by some $r\boldsymbol{t}$ such that $p\boldsymbol{u} \mapsto r\boldsymbol{t}$;
  - if $p \in Q_\diamond$, then $n$ has exactly one successor for every $r\boldsymbol{t}$ such that $p\boldsymbol{u} \mapsto r\boldsymbol{t}$, and the label of this successor is $r\boldsymbol{t}$.
- If $n$ is a leaf of $T$ labeled by $p\boldsymbol{u}$, then there is another node $m$ (where $m \neq n$) on the path from the root of $T$ to $n$ such that the label $p\boldsymbol{t}$ of $m$ satisfies $\boldsymbol{t} < \boldsymbol{u}$ (i.e., $\boldsymbol{t} \leq \boldsymbol{u}$ and $\boldsymbol{t}_\ell < \boldsymbol{u}_\ell$ for at least one index $\ell$).

Consider again a Minsky machine $M$ with two counters $c_1, c_2$ initialized to zero and instructions $1{:}ins_1, \cdots, m{:}ins_m$. We construct a 3-dimensional eVASS game as follows. For every $1 \leq i \leq m$ we add a control state $q_i \in Q_\diamond$. Further, we add a special control state $q_0 \in Q_\square$ and a transition $q_0 \xrightarrow{(0,0,\omega)} q_1$. For every type I instruction $\ell_i : \texttt{inc } c_j;\ \texttt{goto } k$ we add a transition $q_i \to q_k$ labeled by $(u_1, u_2, -1)$, where $u_j = 1$ and $u_1 + u_2 = 1$. For every type II instruction $\texttt{if } c_j{=}0 \texttt{ then goto } k \texttt{ else dec } c_j;\ \texttt{goto } n$ we add control states $p_i, r_i \in Q_\square$, and transitions

$$q_i \xrightarrow{(u_1,u_2,-1)} q_j, \quad q_i \xrightarrow{(0,0,-1)} p_i, \quad p_i \xrightarrow{(0,0,-1)} q_k, \quad p_i \xrightarrow{(u_1,u_2,-1)} r_i, \quad r_i \xrightarrow{(1,1,1)} r_i,$$

where $u_j = -1$ and $u_1 + u_2 = -1$. Finally, we add a transition $q_m \xrightarrow{(1,1,1)} q_m$. Now it is easy to check that $M$ halts iff there is a self-covering tree for $q_0(0, 0, 0)$.

### A.2 A proof of Lemma 7 and Lemma 8

As in Section 3, we fix a $k$-dimensional eVASS game $\mathcal{M} = (Q, (Q_\square, Q_\diamond), T, \alpha, \beta, \delta)$ such that for every transition $p \xrightarrow{\boldsymbol{v}} q$ of $\mathcal{M}$ where $p \in Q_\diamond$ we have that $\boldsymbol{v}_\ell \neq \omega$ for every $\ell \in \{1, 2, \ldots, k\}$. Our aim is to prove the following:

**Lemma 7.** *Let $q(\omega,\ldots,\omega) \in \text{Win}_\square$ (i.e., $q\bm{v} \in \text{Win}_\square$ for some $\bm{v}$). Then there is an unrestricted self-covering tree for $q$ of depth at most $f(|Q|, d, k) = 2^{(d-1)\cdot|Q|} \cdot |Q|^{c\cdot k^2}$, where $c$ is a fixed constant independent of $\mathcal{M}$ (and $d$ is the branching degree of $\mathcal{M}$).*

Lemma 7 is proven in two stages. We start with a special case when $Q_\diamond = \emptyset$. Observe that if $Q_\diamond = \emptyset$, then an (unrestricted) self-covering tree for $q \in Q$ is just a path of the form $q\bm{v} \to^* p\bm{u} \to^+ p\bm{u}'$ where $\bm{v} = (0,\ldots,0)$ and $\bm{u} \leq \bm{u}'$ (recall that $\bm{u}, \bm{u}' \in (\mathbb{Z} \cup \{\omega\})^k$). Below in Lemma 15 we show that if there is *some* path of the above form, then there is also a "*short*" one. The proof is based on arguments similar to the ones used by Rackoff in [28]. However, some extra care is needed to handle the symbolic transitions. Another problem is that the result of [28] is in fact somewhat different, because it studies the existence of an increasing self-covering path for VAS (without states). Therefore, we give an explicit proof.

We then proceed to handle the general case, allowing $Q_\diamond \neq \emptyset$. After a few technical propositions we show Lemma 18, which then easily implies Lemma 7.

We then prove Lemma 19, showing that player $\diamond$ has a counterless winning strategy in each $q(\omega,\ldots,\omega) \in \text{Win}_\diamond$, and finally we derive Lemma 8.

**Lemma 15.** *Assume that $Q_\diamond = \emptyset$ and that $q \in Q$ is a control state such that $q(\omega,\ldots,\omega) \in \text{Win}_\square$. Then there is an unrestricted self-covering tree for $q$ of depth at most $h(|Q|, k) = (|Q| + 1)^{c\cdot k^2}$ where $c$ is a constant independent of $\mathcal{M}$.*

*Proof.* We start by introducing some notation. Let $p, r \in Q$. A *simple sequence from $p$ to $r$* is a sequence of transitions $t_1 \ldots t_n$ such that

- $\alpha(t_1) = p$ and $\beta(t_n) = r$
- for all $1 \leq i < n$ we have $\beta(t_i) = \alpha(t_{i+1})$
- for all $1 \leq i < j \leq n$, where either $i > 1$, or $j < n$, we have $\alpha(t_i) \neq \alpha(t_j)$

A *simple cycle on $p$* is a simple sequence from $p$ to $p$. Given a sequence of transitions $T = t_1 \ldots t_n$, we denote by $e(T)$ the *effect* of $T$ given by $\sum_{i=1}^{n} \delta(t_i)$.

Let $q\bm{v} \to^* p\bm{u} \to^+ p\bm{u}'$ be an unrestricted self-covering tree (a path, in fact) for $q$, where $\bm{u} \leq \bm{u}'$. Obviously, we can safely assume that the sequence of transitions which induces the path $q\bm{v} \to^* p\bm{u}$ is simple (otherwise, we make it simple by repeatedly removing all simple cycles). Let $T = t_1 \ldots t_n$ be the sequence of transitions which induces the path $p\bm{u} \to^+ p\bm{u}'$, and let $q_1, \ldots, q_m$ be *all* control states which occur in transitions of $T$, ordered so that for $i < j$ we have that the first occurrence of $q_i$ precedes the first occurrence of $q_j$ in $T$. For every $1 \leq i \leq m$, we denote by $T_i$ the subsequence $t_j t_{j+1} \ldots t_\ell$ of $T$ where $j$ and $\ell$ are the least indexes such that $\alpha(t_j) = q_i$ and $\beta(t_\ell) = q_{i+1}$, respectively. We also use $T_m$ to denote the (unique) suffix of $T$ such that $T = T_1 \ldots T_{m-1} T_m$.

Now we show that each $T_i$ can be "decomposed" into a simple sequence from $q_i$ to $q_{i+1}$ and a number of simple cycles. Then, we reduce the number of simple cycles needed to obtain an unrestricted self-covering tree for $q$ using similar arguments as in [28].

We start by successively removing simple cycles from $T_i$ (and "remembering" their effects). For every $1 \leq i \leq m$, we construct a sequence of vectors $\bm{w}_i^1, \bm{w}_i^2, \ldots \bm{w}_i^{\xi[i]}$ (where $\xi[i]$ is defined below) and a sequence of transition sequences $T_i^0, T_i^1, \ldots T_i^{\xi[i]}$ as follows:

- $T_i^0 = T_i$
- If $T_i^\ell$ is a simple sequence, set $\xi[i] = \ell$ and stop the construction.

- Otherwise, let $C$ be the first simple cycle in $T_i^\ell$. The sequence $T_i^{\ell+1}$ is obtained from $T_i^\ell$ by removing $C$ and $w_i^{\ell+1}$ is defined to be the effect $e(C)$ of $C$. (Observe that $e(T_i^\ell) = e(T_i^{\ell+1}) + w_i^{\ell+1}$.)

Let $W_i$ be the set $\{w_i^1, w_i^2, \ldots, w_i^{\xi[i]}\}$, and let $W = \bigcup_{i=1}^m W_i$. We have that

$$e(T) = \sum_{i=1}^m \sum_{u \in W_i} n[i, u] \cdot u + e(T_i^{\xi[i]}) = \sum_{u \in W} \left( \sum_{i=1}^m n[i, u] \right) \cdot u + \sum_{i=1}^m e(T_i^{\xi[i]}) \geq 0 \quad (1)$$

where each $n[i, u]$ is the number of occurrences of $u$ in $w_i^1, w_i^2, \ldots w_i^{\xi[i]}$.

When we denote the vector $\sum_{i=1}^m e(T_i^{\xi[i]})$ by $c$ and the sum $\sum_{i=1}^m n[i, u]$ by $n[u]$, the above inequality takes the form

$$\sum_{u \in W} n[u] \cdot u + c \geq 0 \quad (2)$$

Observe that if $n'[u]$ is a non-negative integer for every $u \in W$ and $\sum_{u \in W} n'[u] \cdot u + c \geq 0$, then the tuple of all $n'[u]$ determines a path of the form $pt \to^* pt'$ where $t, t' \in \mathbb{Z}^k$ and $t \leq t'$. To see this, realize that each $u \in W$ is an effect of a simple cycle on some $q_i$, and $c$ is the effect of the sequence $T_1^{\xi[1]} T_2^{\xi[2]} \ldots T_m^{\xi[m]}$. Hence, it suffices to follow the sequence $T_1^{\xi[1]} T_2^{\xi[2]} \ldots T_m^{\xi[m]}$ and whenever a control state $q_i$ is visited for the first time, we do the following: For every $u$ in $W$ which is an effect of a simple cycle $C$ on $q_i$, we perform the cycle $C$ exactly $n'[u]$-times. Whenever $\omega$ occurs in a transition, we set the corresponding counter to a value which is "high enough", i.e., greater than the length of the path we are constructing. Thus, we produce a sequence of transitions with the total effect $\sum_{u \in W} n'[u] \cdot u + c \geq 0$.

Due to the above observations, it suffices to show that there is a tuple $n'[u]$ of "small" non-negative integers such that $\sum_{u \in W} n'[u] \cdot u + c \geq 0$. To achieve that, we use [28, Lemma 4.4] (the lemma was originally proved by Borosh&Treybis [4], but we use the particular form presented in [28]). Since [28, Lemma 4.4] works for systems of equations in real numbers, we have to get rid of $\omega$ components. Note that whenever a transition whose label contains $\omega$ in some component is executed along a path, the corresponding counter can be set to a sufficiently high number to make the path non-decreasing in this particular counter. Hence, we need to make sure that whenever $\omega$ occurs in some component along the original sequence $T$, it also occurs in the same component in the reduced sequence. This is implemented by slightly modifying the system of equations (2) in the way described below.

Let us define
$$\Omega := \{\ell \mid \exists u \in W : u(\ell) = \omega\}$$

To every $u \in W$ we associate a $k$-dimensional vector $u'$ of integers as follows:

$$u'(\ell) := \begin{cases} 0 & \text{if } c(\ell) = \omega; \\ 1 & \text{if } \ell \in \Omega \text{ and } u(\ell) = \omega; \\ 0 & \text{if } \ell \in \Omega \text{ and } u(\ell) \neq \omega; \\ u(\ell) & \text{otherwise.} \end{cases}$$

We define $c'$ by

$$c'(\ell) := \begin{cases} 0 & \text{if } c(\ell) = \omega; \\ -1 & \text{if } \ell \in \Omega; \\ c(\ell) & \text{otherwise.} \end{cases}$$

Note that

$$\sum_{u \in W} n[u] \cdot u' + c' \geq 0$$

On the other hand, an arbitrary tuple of non-negative numbers $n'[u]$ satisfying

$$\sum_{u \in W} n'[u] \cdot u' + c' \geq 0 \tag{3}$$

determines a path $pt \to^* pt'$, where $t \leq t'$ and the length of this path is at most $|Q|^2 + \sum_{u \in W} n'[u] \cdot |Q|$, as follows:

- Start in $q_1 = p$.
- For $i = 1, 2, \ldots, m$ do the following:
  - For every $u \in W$ which is an effect of a simple cycle $C$ on $q_i$ execute the cycle $C$ exactly $n'[u]$-times. Whenever $\omega$ is encountered in some component, add $|Q|^2 + \sum_{u \in W} n'[u] \cdot |Q|$ to the corresponding counter.
  - Follow $T_i^{\xi[i]}$. Whenever $\omega$ is encountered in some component, add $|Q|^2 + \sum_{u \in W} n'[u] \cdot |Q|$ to the corresponding counter.

Now we apply [28, Lemma 4.4] to obtain a small solution of the system of equations (3). First, observe that $|W| \leq (2|Q| + 1)^k$ because each element of $W$ is an effect of a simple cycle. So the number of variables of the system (3), denoted by $d_2$ in [28, Lemma 4.4], is bounded by $(2|Q| + 1)^k$. The absolute values of numbers occurring in the system (3) are bounded by $|Q|^2$ (such numbers may occur only in $c'$, the absolute values of components of $u'$ are bounded by $|Q|$). Thus, $d = (2|Q| + 1)^{2k}$ bounds both $d_2$ and the absolute values of numbers occurring in (3).

By Lemma [28, 4.4], there is a non-negative solution of the system (3) in which all numbers are bounded by $d^{c'k}$ for a suitable constant $c'$ independent of $\mathcal{M}$. Thus there is a constant $c''$, which does not depend on $\mathcal{M}$, such that the absolute values of all numbers occurring in the solution are bounded by $(|Q| + 1)^{c'' \cdot k^2}$. It follows that there is a path $pu \to^* pu''$, $u \leq u''$, whose length is bounded by

$$|Q|^2 + \sum_{u \in W} (|Q| + 1)^{c'' \cdot k^2} \cdot |Q| = (|Q| + 1)^{c \cdot k^2}$$

for a suitable constant $c$ independent of $\mathcal{M}$. □

The following proposition is crucial for handling the general case (where $Q_\diamond$ can be non-empty).

**Proposition 16.** *Suppose $q' \in Q_\diamond$ in $\mathcal{M}$ has more than one outgoing transition, namely an outgoing transition $t$ and a nonempty set $R$ (the 'rest') of other outgoing transitions; by $\mathcal{M}_1$ ($\mathcal{M}_2$) we denote the eVASS arising from $\mathcal{M}$ by removing $R$ (t). Suppose now that, for some state $q$ which may be different from $q'$, we have $q\mathbf{v_1} \in Win_\square$ in $\mathcal{M}_1$ and $q'\mathbf{v_2} \in Win_\square$ in $\mathcal{M}_2$. Then $q(\mathbf{v_1}+\mathbf{v_2}-\mathbf{1}) \in Win_\square$ in $\mathcal{M}$.*

*Proof.* Let $S_1$ be a winning strategy of player $\square$ in $q\mathbf{v_1}$ in $\mathcal{M}_1$, and $S_2$ a winning strategy of player $\square$ in $q'\mathbf{v_2}$ in $\mathcal{M}_2$. The following strategy will be winning for player $\square$ in $q(\mathbf{v_1}+\mathbf{v_2}-\mathbf{1})$ in $\mathcal{M}$:

Player $\square$ uses the strategy $S_1$ as long as player $\diamond$ does not use any transition from the set $R$ (when the play goes through $q'$). If this happens, i.e. player $\diamond$ uses some $t' \in R$, then player $\square$ suspends the strategy $S_1$ and behaves according to $S_2$ (starting in $q'$). If player $\diamond$ uses $t$ in future, player $\square$ just suspends $S_2$ and resumes the (previously suspended) $S_1$, etc.

Thus every prefix of any play arises by merging two prefixes of particular plays, played from $q\mathbf{v_1}$ in $\mathcal{M}_1$ according to $S_1$ and from $q'\mathbf{v_2}$ in $\mathcal{M}_2$ according to $S_2$, respectively. Any prefix of the first (second) particular play cannot decrease a counter $j$ by more than $(\mathbf{v_1})_j-1$ $((\mathbf{v_2})_j-1)$, and thus their merging keeps the value of each counter above zero, when starting from $\mathbf{v_1}+\mathbf{v_2}-\mathbf{1}$. □

The following simple proposition is technically useful.

**Proposition 17.** *Assume an unrestricted self-covering tree $\mathcal{T}$ and a leaf (labelled with) $p\mathbf{u'}$, having above a corresponding node $p\mathbf{u}$ with $\mathbf{u} \leq \mathbf{u'}$. Consider the tree $\mathcal{T}'$ arising by an "unfolding", i.e., arising by hanging a corresponding copy of the (original) subtree rooted in $p\mathbf{u}$ on the node $p\mathbf{u'}$. (Each node labelled with $r\mathbf{t}$ in the original subtree has a corresponding node in the newly hanged subtree, labelled with $r\mathbf{t'}$ where $\mathbf{t'} = \mathbf{t}+\mathbf{u'}-\mathbf{u}$.) Then $\mathcal{T}'$ is also an unrestricted self-covering tree.*

We now want to generalize Lemma 15. We first note that the case when player $\diamond$ has no choice, i.e. when the set $tr(Q_\diamond)$ of transitions $t$ with $\alpha(t) \in Q_\diamond$ has the same cardinality as $Q_\diamond$ (recall that each control state has at least one outgoing transition), is already handled by Lemma 15: in such a case, all states in $Q_\diamond$ can be viewed as being in $Q_\square$, in fact.

In the general case we take the number $r = |tr(Q_\diamond)| - |Q_\diamond|$ as a suitable measure of the *choice degree of $\diamond$*.

**Lemma 18.** *(Given eVASS $\mathcal{M}$), let $q \in Q$ be a control state such that $q\mathbf{v} \in Win_\square$ for some $\mathbf{v}$. Then there is an unrestricted self-covering tree for $q$ of depth at most $2^r \cdot h(|Q|, k)$ where $h$ is the function from Lemma 15 and $r$ is the choice degree of $\diamond$, i.e. the number $|tr(Q_\diamond)| - |Q_\diamond|$.*

*Proof.* We proceed by induction on $r$. The base case $r = 0$ has been already handled, so we assume the claim holds for $r$, and show it for $r + 1$. Let $q' \in Q_\diamond$ be a fixed state with at least two choices, i.e., with an outgoing transition $t$ and a nonempty set $R$ (the 'rest') of other outgoing transitions.

Let $\mathcal{M}_1$ be the eVASS arising from $\mathcal{M}$ by removing $R$, and let $\mathcal{M}_2$ be the eVASS arising from $\mathcal{M}$ by removing $t$; the choice degree of $\diamond$ is at most $r$ in both $\mathcal{M}_1$ and $\mathcal{M}_2$.

Let us now consider a control state $q$ such that $q\mathbf{v} \in Win_\square$ for some $\mathbf{v}$ in $\mathcal{M}$; obviously, $q\mathbf{v} \in Win_\square$ in both $\mathcal{M}_1$ and $\mathcal{M}_2$ as well. If some of the unrestricted self-covering trees of depth at most $2^r \cdot h(|Q|, k)$ which are guaranteed by the induction hypothesis does not contain $q'$, then we are done. If both of them contain $q'$ then $q'$ must have unrestricted self-covering trees in both $\mathcal{M}_1$ and $\mathcal{M}_2$ (recall Proposition 17); in particular, $q'\mathbf{v'} \in Win_\square$ for some $\mathbf{v'}$ in $\mathcal{M}_2$, and by the induction hypothesis the elements of $\mathbf{v'}$ do not need to exceed $2^r \cdot h(|Q|, k)$. The rest follows from Proposition 16. □

We now define $Q^{D-inf} = \{q \in Q \mid q(\omega, \ldots, \omega) \in Win_\diamond\}$ and show that there is a fixed counter-less strategy of player $\diamond$ which is winning inside "$Q^{D-inf}$-area".

**Lemma 19.** *There is a counter-less strategy of player $\diamond$ which is winning in every $q\mathbf{v} \in Q^{D-inf} \times \mathbb{N}^k$ and, moreover, all control states visited in the respective plays are in $Q^{D-inf}$.*

*Proof.* We proceed by induction on the size of the underlying eVASS $\mathcal{M}$; in other words, we assume that the claim holds for all eVASSs with lesser sizes than the size of $\mathcal{M}$ and we prove the claim for $\mathcal{M}$.

We observe that there is no transition $t : q \to q'$ such that $q \in Q^{D-inf} \cap Q_\square$ and $q' \notin Q^{D-inf}$ (otherwise $q'\mathbf{v}' \in Win_\square$ for some $\mathbf{v}'$ and thus $q\mathbf{v} \in Win_\square$ for some $\mathbf{v}$ which contradicts with $q(\omega, \ldots, \omega) \in Win_\diamond$). We can also easily verify that each $q \in Q^{D-inf} \cap Q_\diamond$ has at least one outgoing transition leading to $Q^{D-inf}$; if there is, moreover, some $t : q \to q'$ such that $q' \notin Q^{D-inf}$, then removing $t$ results in a lesser $\mathcal{M}'$ with the same $Q^{D-inf}$ and the claim for $\mathcal{M}$ follows by the induction hypothesis. (If $b$ is the maximal component in $Min_\square$ then $q'(b, \ldots, b) \in Win_\square$ for all $q' \notin Q^{D-inf}$. Starting from $q\mathbf{v}$, $q \in Q^{D-inf}$, player $\diamond$ can use the same strategy as from $q(\mathbf{v} + (b, \ldots, b))$, thus reaching 0 in some counter without leaving the $Q^{D-inf}$-area.)

It thus remains to explore the case with no transitions leaving $Q^{D-inf}$; moreover $Q^{D-inf} = Q$ since otherwise we also finish by the induction hypothesis. If now $\mathcal{M}$ has only one outgoing transition for every $q \in Q_\diamond$ then the claim is obvious, so we assume that at least one $q \in Q_\diamond$ in $\mathcal{M}$ has more than one outgoing transition, namely an outgoing transition $t$ and a nonempty set $R$ (the 'rest') of other outgoing transitions.

We define $\mathcal{M}_1, \mathcal{M}_2$ as in Proposition 16; and we start with assuming that $q \in Q^{D-inf}$ also in $\mathcal{M}_1$. Then $Q^{D-inf}$ in $\mathcal{M}_1$ coincides with $Q^{D-inf} = Q$ in $\mathcal{M}$ (and the claim thus follows from the induction hypothesis): from every $q'\mathbf{v}'$ in $\mathcal{M}_1$, player $\diamond$ can use his winning strategy $S$ from $q'\mathbf{v}'$ in $\mathcal{M}$ until (winning or) possibly reaching some $q\mathbf{v}$ where $S$ prescribes to use some $t' \in R$; here player $\diamond$ switches to his winning strategy which is guaranteed by the assumption that $q \in Q^{D-inf}$ in $\mathcal{M}_1$. Similarly we handle the case when $q \in Q^{D-inf}$ in $\mathcal{M}_2$.

Thus it remains to consider the case when $q \notin Q^{D-inf}$ in $\mathcal{M}_1$ nor in $\mathcal{M}_2$; then necessarily $q\mathbf{v_1} \in Win_\square$ in $\mathcal{M}_1$ and $q\mathbf{v_2} \in Win_\square$ in $\mathcal{M}_2$ for some $\mathbf{v_1}, \mathbf{v_2} \in \mathbb{N}^k$. But then Proposition 16 yields a contradiction with the assumption $q \in Q^{D-inf}$ in $\mathcal{M}$. □

**Lemma 8.** *The set of all $q \in Q$ such that $q(\omega, \ldots, \omega) \in Win_\diamond$ is computable in space bounded by a polynomial function $g(|Q|, d, k)$.*

*Proof.* It is sufficient to check (successively, in the same working space) all counterless strategies of player $\diamond$ (recall Lemma 19); each case amounts to prune some outgoing transitions for each $q \in Q_\diamond$ so that always just one is left.

Now to show that a particular counterless strategy is *not* winning for player $\diamond$ in some $q\mathbf{v}$, i.e. that $q(\omega, \ldots, \omega) \in Win_\square$ in the pruned system, we recall Lemma 15 (i.e., the case $r = 0$ of Lemma 18). We can nondeterministically go along a self-covering path, remembering only the current configuration and the beginning of the cycle after we guess we have encountered it. Polynomial space is obviously sufficient for this procedure.

□

### A.3 A proof of Lemma 11

We start with the following auxiliary observation:

**Lemma 20.** *Let $0 \leq j < k$, and let $B \in \mathbb{N}$ be a bound which is safe for precision $j$. Then for every $q\mathbf{v} \in SymMin_\square^{j+1}$ we have that $\mathbf{v}_\ell \leq B$ for every $\ell$ precise in $q\mathbf{v}$.*

*Proof.* Let $q\mathbf{v} \in \mathit{SymMin}_\square^{j+1}$, and let us assume that $\mathbf{v}_\ell > B$ for some $\ell$ precise in $q\mathbf{v}$. Let $q\mathbf{u}$ be a symbolic configuration where $\mathbf{u}_\ell = \mathbf{v}_\ell$ for all $\ell$ such that $\mathbf{v}_\ell \leq B$, and $\mathbf{u}_\ell = \omega$ for the other $\ell$. Note that $P(q\mathbf{u}) \leq j$, and $q\mathbf{u} \in \mathit{Win}_\square$ because $q\mathbf{v} \in \mathit{Win}_\square$. Hence, there is some $q\mathbf{t} \in \bigcup_{i=0}^{j} \mathit{SymMin}_\square^i$ such that $q\mathbf{t} \leq q\mathbf{v}$. Since $B$ is safe for precision $j$, we have that $q\mathbf{t}' \in \mathit{Win}_\square$, where $\mathbf{t}'$ is obtained from $\mathbf{t}$ by replacing every $\omega$-component with $B$. Since $q\mathbf{t}' \leq q\mathbf{v}$ and $\mathbf{t}'_\ell < \mathbf{v}_\ell$ for at least one $\ell$ precise in $q\mathbf{v}$, we obtain a contradiction with the minimality of $q\mathbf{v}$. □

Now we have all the tools needed to prove Lemma 11.

**Lemma 11.** *Let $0 \leq j < k$, and let us assume that $\bigcup_{i=0}^{j} \mathit{SymMin}_\square^i$ has already been computed, together with some bound $B \in \mathbb{N}$ which is safe for precision $j$. Then $\mathit{SymMin}_\square^{j+1}$ is computable in time exponential in $|Q| \cdot B^{j+1}$, $d$, and $k-j-1$, and the bound $B + f(|Q| \cdot B^{j+1}, d, k-j-1)$ is safe for precision $j+1$ (here $f$ is the function of Lemma 7 and $d$ is the branching degree of $\mathcal{M}$).*

*Proof.* Let us fix some subset $C$ of $\{1, \ldots, k\}$ of cardinality $j+1$, and let $\bar{C} = \{1, \ldots, k\} \setminus C$. We show how to compute the set of all $q\mathbf{v} \in \mathit{SymMin}_\square^{j+1}$ such that the set of all indexes that are precise in $q\mathbf{v}$ is exactly $C$. To achieve that, we construct an alternating eVASS $\mathcal{M}_C$ with $k - j - 1$ counters which encodes the counter values indexed by the elements of $C$ in its finite control (up to the bound $B$) and simulates the execution of the considered eVASS $\mathcal{M}$. Hence, the counters of $\mathcal{M}_C$ simulate the counters of $\mathcal{M}$ that are indexed by the elements of $\bar{C}$. For every configuration $p\mathbf{x}$ of $\mathcal{M}_C$ and every $\ell \in \bar{C}$, we use $\mathbf{x}_\ell$ to denote the current value of the counter which corresponds to the $\ell$-th counter of $\mathcal{M}$. Similarly, if $\mathbf{y}$ is a tuple of counter changes in $\mathcal{M}_C$ and $\ell \in \bar{C}$, we use $\mathbf{y}_\ell$ to denote the change on the counter of $\mathcal{M}_C$ which corresponds to the $\ell$-th counter of $\mathcal{M}$. This convention leads to a simpler notation.

The simulation of $\mathcal{M}$ by $\mathcal{M}_C$ is essentially faithful until the point when some of the counters indexed by $C$ either reaches zero or attempts to cross the bound $B$. In the first case, $\mathcal{M}_C$ enters a special control state where player $\diamond$ wins (for arbitrary counter values). In the latter case, the behaviour of $\mathcal{M}_C$ is more subtle and it is explained later.

The set of control states of $\mathcal{M}_C$ consists of $q_\square$, $q_\diamond$, and all elements of $Q \times (C \to \{1, \ldots, B\})$. The states of $Q_\diamond \times (C \to \{1, \ldots, B\})$ belong to player $\diamond$, and the other states belong to player $\square$. To each control state of the form $(p, \mathbf{a})$ we associate the (unique) symbolic configuration $p[\mathbf{a}]$ of $\mathcal{M}$ where $[\mathbf{a}]_\ell = \mathbf{a}_\ell$ for all $\ell \in C$ such that $\mathbf{a}_\ell < B$, and $[\mathbf{a}]_\ell = \omega$ for all of the remaining indexes $\ell$. The transitions of $\mathcal{M}_C$ together with their labels are constructed as follows:

- There is a transition $q_\square \to q_\square$ labeled by $(0, \ldots, 0)$ and a transition $q_\diamond \to q_\diamond$ labeled by $(-1, \ldots, -1)$.
- For every transition $p \xrightarrow{\mathbf{u}} q$ of $\mathcal{M}$ we add the following transitions to $\mathcal{M}_C$:
  (a) For all $\mathbf{a} : C \to \{1, \ldots, B\}$ such that $\mathbf{u}_\ell = -1$ and $\mathbf{a}_\ell = 1$ for some $\ell \in C$, we add a transition $(p, \mathbf{a}) \to q_\diamond$ labeled by $(0, \ldots, 0)$.
  (b) For all $\mathbf{a} : C \to \{1, \ldots, B\}$ such that the previous item does not apply and $\mathbf{a}_\ell < B$ for all $\ell \in C$, we add a transition $(p, \mathbf{a}) \xrightarrow{\mathbf{x}} (q, \mathbf{b})$, where $\mathbf{b}$ and $\mathbf{x}$ are the unique vectors satisfying the following:
    * For every $\ell \in C$ we have that $\mathbf{b}_\ell$ is equal either to $\mathbf{a}_\ell + \mathbf{u}_\ell$ or $B$, depending on whether $\mathbf{u}_\ell \in \{-1, 0, 1\}$ or $\mathbf{u}_\ell = \omega$, respectively.
    * For all $\ell \in \bar{C}$ we have that $\mathbf{x}_\ell = \mathbf{u}_\ell$.

(c) For all $\boldsymbol{a} : C \to \{1,\ldots,B\}$ such that $\boldsymbol{a}_\ell = B$ for some $\ell \in C$, we add either a transition $(p,\boldsymbol{a}) \to q_\square$ labeled by $(0,\ldots,0)$ or a transition $(p,\boldsymbol{a}) \to q_\diamond$ labeled by $(0,\ldots,0)$, depending on whether $p[\boldsymbol{a}] \in Win_\square$ or $p[\boldsymbol{a}] \in Win_\diamond$, respectively. Realize that since $P(p[\boldsymbol{a}]) \leq j$, we have that $p[\boldsymbol{a}] \in Win_\square$ iff there is $p\boldsymbol{v} \in \bigcup_{i=0}^{j} SymMin_\square^i$ such that $p\boldsymbol{v} \leq p[\boldsymbol{a}]$, which can be checked effectively because the set $\bigcup_{i=0}^{j} SymMin_\square^i$ has already been computed.

In the rest of this proof, the winning regions for player $\square$ and player $\diamond$ in $G_{\mathcal{M}_C}$ are denoted by $Win_\square(\mathcal{M}_C)$ and $Win_\diamond(\mathcal{M}_C)$, respectively. For a given symbolic configuration $(p,\boldsymbol{a})\boldsymbol{x}$ of $\mathcal{M}_C$, we use $p(\boldsymbol{a},\boldsymbol{x})$ to denote the corresponding symbolic configuration of $\mathcal{M}$, i.e., $(\boldsymbol{a},\boldsymbol{x})_\ell = \boldsymbol{a}_\ell$ for all $\ell \in C$, and $(\boldsymbol{a},\boldsymbol{x})_\ell = \boldsymbol{x}_\ell$ for all $\ell \in \bar{C}$. For the moment, assume that the following two claims are already proven (where $f$ is the function of Lemma 7):

(1) If $(p,\boldsymbol{a})\boldsymbol{x} \in Win_\square(\mathcal{M}_C)$ where $\boldsymbol{x}_\ell \geq B + f(|Q| \cdot B^{j+1}, k-j-1)$ for every $\ell \in \bar{C}$, then $p(\boldsymbol{a},\boldsymbol{x}) \in Win_\square$.
(2) If $(p,\boldsymbol{a})\boldsymbol{x} \in Win_\diamond(\mathcal{M}_C)$, then $p(\boldsymbol{a},\boldsymbol{x}) \in Win_\diamond$.

An immediate consequence of (1) and (2) is that $(p,\boldsymbol{a})(\omega,\ldots,\omega) \in Win_\square(\mathcal{M}_C)$ iff $p(\boldsymbol{a},\omega,\ldots,\omega) \in Win_\square$. By applying Lemma 20, it follows that $SymMin_\square^{j+1}$ contains exactly the minimal $p(\boldsymbol{a},\omega,\ldots,\omega)$ such that $(p,\boldsymbol{a})(\omega,\ldots,\omega) \in Win_\square(\mathcal{M}_C)$. Since the set of all $(p,\boldsymbol{a})$ such that $(p,\boldsymbol{a})(\omega,\ldots,\omega) \in Win_\square(\mathcal{M}_C)$ is computable in time exponential in $|Q| \cdot B^{j+1}$, $d$, and $k-j-1$ by Lemma 8, the set $SymMin_\square^{j+1}$ is also computable in time exponential in $|Q| \cdot B^{j+1}$, $d$, and $k-j-1$ (although the number of control states of $\mathcal{M}_C$ is actually $(|Q| \cdot B^{j+1}) + 2$, note that we can easily adjust $\mathcal{M}_C$ by removing the control states $q_\diamond$ and $q_\square$ without influencing the winning regions for the other control states). Moreover, the bound $B + f(|Q| \cdot B^{j+1}, k-j-1)$ is obviously safe for precision $j+1$. So, it remains to prove Claims (1) and (2). First, recall that the transitions of $\mathcal{M}_C$ introduced in item (c) above are "correct" in the following sense: if $(p,\boldsymbol{a})$ is a control state of $\mathcal{M}_C$ such that $\boldsymbol{a}_\ell = B$ for some $\ell \in C$, then

– if $(p,\boldsymbol{a}) \to q_\square$, then $p(\boldsymbol{a},\boldsymbol{y}) \in Win_\square$ for all $\boldsymbol{y}$ such that $\boldsymbol{y}_\ell \geq B$ for every $\ell \in \bar{C}$;
– if $(p,\boldsymbol{a}) \to q_\diamond$, then $p(\boldsymbol{a},\boldsymbol{y}) \in Win_\diamond$ for all $\boldsymbol{y}$.

Claim (1): Let us assume that $(p,\boldsymbol{a})\boldsymbol{x} \in Win_\square(\mathcal{M}_C)$ where $\boldsymbol{x}_\ell \geq B + f(|Q| \cdot B^{j+1}, k-j-1)$. Since $(p,\boldsymbol{a})\boldsymbol{x} \in Win_\square(\mathcal{M}_C)$, by Lemma 7 there is a self-covering tree $T$ for $(p,\boldsymbol{a})$ of depth at most $f(|Q| \cdot B^{j+1}, k-j-1)$. A winning strategy for player $\square$ in $p(\boldsymbol{a},\boldsymbol{x})$ is obtained simply by following the strategy described by $T$ until the point when a transition of the form $(q,\boldsymbol{b})\boldsymbol{z} \to q_\square \boldsymbol{z}$ is to be executed in $T$. Note that since the depth of $T$ is at most $f(|Q| \cdot B^{j+1}, k-j-1)$, we have that $(\boldsymbol{x}+\boldsymbol{z})_\ell \geq B$ for every $\ell \in \bar{C}$. This means that $q(\boldsymbol{b},\boldsymbol{x}+\boldsymbol{z}) \in Win_\square$ (see above) and hence player $\square$ can simply abandon the strategy described by $T$ and start to follow his winning strategy for $q(\boldsymbol{b},\boldsymbol{x}+\boldsymbol{z})$.

Claim (2): Let us assume that $(p,\boldsymbol{a})\boldsymbol{x} \in Win_\diamond(\mathcal{M}_C)$. A winning strategy for player $\diamond$ in $p(\boldsymbol{a},\boldsymbol{x})$ is obtained simply by "mimicking" the winning strategy of player $\diamond$ in $(p,\boldsymbol{a})\boldsymbol{x}$ until one of the two players enters a configuration $(q,\boldsymbol{b})\boldsymbol{y}$ which has an outgoing transition of the form $(q,\boldsymbol{b})\boldsymbol{y} \to q_\diamond \boldsymbol{y}$. Note that then there must be at least one outgoing transition of $q(\boldsymbol{b},\boldsymbol{y})$ leading to a winning configuration of player $\diamond$. If $q \in Q_\diamond$, then player $\diamond$ selects this transition and "switches" to the winning strategy of the chosen successor. If $q \in Q_\square$, then player $\square$ may select an outgoing transition of $q(\boldsymbol{b},\boldsymbol{y})$ which either does or does not correspond to the transition $(q,\boldsymbol{b})\boldsymbol{y} \to q_\diamond \boldsymbol{y}$ (see item (c) above). In the first case, player $\diamond$ "switches" to the winning strategy for the chosen successor, and in the latter case he keeps

"mimicking" the winning strategy for $(p, a)x$. Obviously, if player $\diamond$ plays in the way just described, he has to win. □

An immediate corollary to Lemma 11 is that the set $SymMin_\square$ (and hence also the set $Min_\square$) is effectively computable (an upper complexity bound is given in Theorem 12). Let us note that the set $Max_\diamond$ of all maximal symbolic configurations which belong to $Win_\diamond$ is effectively computable (we just need to complement the upward closure of $Min_\square$, which can be done by standard methods; see, e.g., [33]).

Finally, we show that there is a finitely and effectively representable strategy $\pi$ of player $\diamond$ which is winning in every configuration of $Win_\diamond$. Let $C$ be a subset of $\{1, \ldots, k\}$, $\bar{C} = \{1, \ldots, k\} \smallsetminus C$, and let

$$Max_\diamond^C = \{p\boldsymbol{v} \in Max_\diamond \mid P(p\boldsymbol{v}) = |C| \text{ and } \boldsymbol{v}_\ell \neq \omega \text{ for all } \ell \in C\}.$$

We also use $\downarrow Max_\diamond^C$ to denote the *downwards closure* of $Max_\diamond^C$, i.e., the set of all configurations $q\boldsymbol{u}$ where $q\boldsymbol{u} \leq q\boldsymbol{u}'$ for some $q\boldsymbol{u}' \in Max_\diamond^C$. The *C-part* of a configuration $q\boldsymbol{u} \in \downarrow Max_\diamond^C$ is a pair $q[\boldsymbol{u}, C]$ where $[\boldsymbol{u}, C] : C \to \mathbb{N}$ such that $[\boldsymbol{u}, C]_\ell = \boldsymbol{u}_\ell$ for every $\ell \in C$. Note that the set

$$Adm_C = \{q[\boldsymbol{u}, C] \mid q\boldsymbol{u} \in \downarrow Max_\diamond^C\}$$

of *admissible C-parts* is finite.

Let $B$ a bound which is safe for precision $k$ (see Lemma 11). Let $q\boldsymbol{u} \in Win_\diamond$ be a configuration such that $q[\boldsymbol{u}, C] \notin Adm_C$ and $\boldsymbol{u}_\ell \geq B$ for every $\ell \in C$. Then there must a *proper* subset $C'$ of $C$ such that $q[\boldsymbol{u}, C'] \in Adm_{C'}$ (otherwise, $q\boldsymbol{u} \in Win_\square$ which is a contradiction). We show that there is a memoryless strategy $\pi_C$ for player $\diamond$ with the following properties:

- For every $p\boldsymbol{v} \in Max_\diamond^C$ and every strategy $\sigma$ of player $\square$ we have that the play initiated in $p\boldsymbol{v}$ reaches either a configuration of $Z$ or a configuration $q\boldsymbol{u}$ such that $q[\boldsymbol{u}, C'] \in Adm_{C'}$ for some proper subset $C'$ of $C$.
- For every $p\boldsymbol{v}$ such that $p \in Q_\diamond$ we have that $\pi_C(p\boldsymbol{v})$ depends only on the $C$-part of $p\boldsymbol{v}$.

Observe that the strategies $\pi_C$ can be easily combined into the promised strategy $\pi$, which works in the following way: for a given configuration $p\boldsymbol{v} \in Win_\diamond$, we find a minimal $C \subseteq \{1, \ldots, k\}$ such that $p\boldsymbol{v} \in Max_\diamond^C$ (if there are more candidates for $C$, any of them can be chosen in some deterministic fashion). Player $\diamond$ plays according to $\pi_C$ untill he either wins or enters a configuration $q\boldsymbol{u}$ such that $q[\boldsymbol{u}, C'] \in Adm_{C'}$ for some proper subset $C'$ of $C$. From this point on, he "switches" to $\pi_{C'}$. Note that such a "switch" can be performed at most $k$ times in each play, and the strategy $\pi$ admits a finite and effective description.

So, it remains to show how to construct the strategy $\pi_C$. Note that if $C = \emptyset$, then $\pi_C$ is counterless by Lemma 19 and can be constructed effectively. Otherwise, we proceed similarly as in Lemma 11. We construct another eVASS game $\mathcal{M}_C$ with $k - C$ counters which simulates the $C$-parts of configurations in its finite control so that

- the transitions of $\mathcal{M}$ that would lead to configurations $q\boldsymbol{u}$ such that $q[\boldsymbol{u}, C'] \in Adm_{C'}$ for some proper subset $C'$ of $C$ are simulated by entering a special control state where player $\diamond$ wins;
- the transitions of $\mathcal{M}$ that would lead to configurations $q\boldsymbol{u}$ such that $q[\boldsymbol{u}, C'] \notin Adm_{C'}$ for every $C' \subseteq C$ are simulated by entering a special control state where player $\square$ wins.

We determine all control states of $\mathcal{M}_C$ such that player $\diamond$ wins for all values in the $k - C$ counters of $\mathcal{M}_C$ and construct the corresponding counterless winning strategy (here we again relay on Lemma 8). Then we "transfer" this counterless strategy back to $\mathcal{M}$ and produce the desired $\pi_C$.

### A.4 Proofs of Theorem 13 and Theorem 14

For the rest of this section, we fix a one-dimensional eVASS game $\mathcal{M} = (Q, (Q_\square, Q_\diamond), T, \alpha, \beta, \delta)$ and $C \subseteq Q$. Recall that for every $i \in \mathbb{N}$, we use $Win_\diamond(C, i)$ to denote the set $\{p \in Q \mid p(i) \in Win(\diamond, \mathcal{R}(Z_C))\}$. Observe that if $Win_\diamond(C, i) = Win_\diamond(C, j)$ for some $i, j \in \mathbb{N}$, then also $Win_\diamond(C, i+1) = Win_\diamond(C, j+1)$. To see this, realize that if $p(i+1) \in Win(\diamond, \mathcal{R}(Z_C))$, then player $\diamond$ has a winning strategy $\pi$ in $p(i+1)$ and hence he can enforce descreasing the counter to $i$ (and entering some configuration $q(i)$ where $q \in Win_\diamond(C, i)$) no matter what player $\square$ does. Then a strategy $\pi^{+(j-i)}$ such that $\pi^{+(j-i)}(r(k)) = \pi(r(k-j+i))$ for all $k \geq j$ can be used in $p(j+1)$ to enforce visiting a configuration $q(j)$ where $q \in Win_\diamond(C, j)$. Since $q(j) \in Win(\diamond, \mathcal{R}(Z_C))$, we obtain that $p(j+1) \in Win(\diamond, \mathcal{R}(Z_C))$. Similarly, one can show that if $p(j+1) \in Win(\diamond, \mathcal{R}(Z_C))$, then also $p(i+1) \in Win(\diamond, \mathcal{R}(Z_C))$. Further, recall that we use $m_C$ to denote the least $i \in \mathbb{N}$ such that $Win_\diamond(C, i) = Win_\diamond(C, j)$ for some $j > i$, and $n_C$ to denote the least $i > 0$ such that $Win_\diamond(C, m_C) = Win_\diamond(C, m_C+i)$. Observe that $m_c + n_c \leq 2^{|Q|}$, and for every $i \geq m_c$ we have that $Win_\diamond(C, i) = Win_\diamond(C, m_C + ((i - m_C) \mod n_C))$. Hence, the winning regions of both players are fully characterized by all $Win_\diamond(C, i)$, where $0 \leq i < m_C + n_C$.

We start with the non-selective case, because some of the underlying observations are needed to solve the more general selective case. Recall that in the non-selective case, $m_Q \leq |Q|$ and $n_Q = 1$, because $Win_\diamond(Q, i) \supseteq Win_\diamond(Q, i+1)$ for every $i \in \mathbb{N}$. Hence, it suffices to compute all $Win_\diamond(Q, i)$ where $0 \leq i \leq |Q|$. The next lemma says that this can be done in polynomial time.

**Lemma 21.** *The sets $Win_\diamond(Q, i)$, where $0 \leq i \leq |Q|$, are computable in $O(|\mathcal{M}|^2)$ time.*

*Proof.* Let $\mathcal{D}$ be the domain of all $|Q|+1$-tuples of subsets of $Q$, ordered by componentwise inclusion. For a given $D \in \mathcal{D}$, the individual components of $D$ are denoted by $D_0, \ldots, D_{|Q|}$. The least element of $\mathcal{D}$ (i.e., the tuple of empty sets) is denoted by $\bot$.

The algorithm computes the least fixed-point of the function $\mathcal{F} : \mathcal{D} \to \mathcal{D}$ defined as follows: $p \in (\mathcal{F}(D))_i$ iff one of the following conditions holds:

- $i = 0$;
- $i > 0$, $p \in Q_\diamond$, and there is an edge $p(i) \mapsto q(j)$ such that either $j \leq |Q|$ and $q \in D_j$, or $j > |Q|$ and $q \in D_{|Q|}$;
- $i > 0$, $p \in Q_\square$, and for every edge $p(i) \mapsto q(j)$ we have that either $j \leq |Q|$ and $q \in D_j$, or $j > |Q|$ and $q \in D_{|Q|}$.

Since $\mathcal{F}$ is continuous, the least fixed-point of $\mathcal{F}$ is equal to $\bigcup_{k \in \mathbb{N}} \mathcal{F}^k(\bot)$, where $\cup$ is considered componentwise. We claim that

$$\bigcup_{k \in \mathbb{N}} \mathcal{F}^k(\bot) = \left(Win_\diamond(Q, 0), \ldots, Win_\diamond(Q, |Q|)\right)$$

The "$\subseteq$" is proven by a straightforward induction on $k$. Note that since $m_Q \leq |Q|$ and $n_Q = 1$, we have that $Win_\diamond(Q, |Q|) = Win_\diamond(Q, |Q|+1)$, and this fact is used to justify the case when $p(i)$ performs an edge which increases the counter above $|Q|$. For the "⊇" direction, consider the set $B$ of configurations defined as follows: $p(i) \in B$ iff $p \notin (\mathcal{F}^k(\bot))_\ell$ for all $k \in \mathbb{N}$, where $\ell = \min\{i, |Q|\}$. We show that $B \subseteq Win(\square, \mathcal{S}(Z))$. To see this, realize the following:

- if $p(i) \in B$ and $p \in Q_\diamond$, then for every edge $p(i) \mapsto q(j)$ we have that $q(j) \in B$;
- if $p(i) \in B$ and $p \in Q_\square$, then there is an edge $p(i) \mapsto q(j)$ such that $q(j) \in B$.

Both claims follow directly from the definition of $\mathcal{F}$. Hence, we can setup a strategy $\sigma \in \Sigma$ which is $\mathcal{S}(Z)$-winning for player $\square$ in every configuration of $B$, which means that $B \subseteq Win(\square, \mathcal{S}(Z))$. From this we obtain that if $p \notin (\mathcal{F}^k(\bot))_i$ for every $k \in \mathbb{N}$, then $p(i) \in B \subseteq Win(\square, \mathcal{S}(Z))$, which means $p \notin Win_\diamond(Q, i)$.

It remains to show that $\bigcup_{k \in \mathbb{N}} \mathcal{F}^k(\bot)$ is computable in polynomial time (this is not completely trivial, because $\mathcal{D}$ has $2^{O(|Q|^2)}$ elements). We say that $D \in \mathcal{D}$ is *monotone* if $D_i \supseteq D_{i+1}$ for all $0 \leq i < |Q|$. Observe that $\bot$ is monotone, and if $D$ is monotone then $\mathcal{F}(D)$ is monotone. Since the length of every increasing chain $D^0 \supset D^1 \supset D^2 \cdots$ where all $D^j \in \mathcal{D}$ are monotone is bounded by $|Q| \cdot (|Q|+1)$, we have that $\bigcup_{k \in \mathbb{N}} \mathcal{F}^k(\bot) = \bigcup_{k=1}^{|Q| \cdot (|Q|+1)} \mathcal{F}^k(\bot)$ and hence the least fixed point of $\mathcal{F}$ is computable in $O(|\mathcal{M}|^2)$ time. □

According to Lemma 21, the problem whether $p(i) \in Win(\diamond, \mathcal{R}(Z))$ for a given configuration $p(i)$ of $\mathcal{M}$ is in **P**, and a finite description of the winning regions for both players is computable in polynomial time. Our next lemma reveals that both players have fixed counterless strategies computable in polynomial time that are winning in every configuration of the corresponding winning region.

**Lemma 22.** *There are counterless strategies $\hat{\pi}$ and $\hat{\sigma}$ computable in polynomial time such that $\hat{\pi}$ is $\mathcal{R}(Z)$-winning for player $\diamond$ in every configuration of $Win(\diamond, \mathcal{R}(Z))$, and $\hat{\sigma}$ is $\mathcal{S}(Z)$-winning for player $\square$ in every configuration of $Win(\square, \mathcal{S}(Z))$.*

*Proof.* The construction of $\hat{\sigma}$ is simple. We just need to ensure that player $\square$ never leaves his winning region. For every $p \in Q_\square$, we fix a transition $t_p \in T$ where $\alpha(t_p) = p$ as follows:

- if $p \in Win_\diamond(Q, |Q|)$, then $t_p$ is chosen arbitrarily;
- otherwise, let $i \in \mathbb{N}$ be the least index such that $p \notin Win_\diamond(Q, i)$. According to the proof of Lemma 21, there is an edge $p(i) \mapsto q(j)$ such that $q(j) \notin Win_\diamond(Q, j)$. We choose $t_p$ to be the transition which induces the edge $p(i) \mapsto q(j)$ (if there are more candidates for $t_p$, any of them can be chosen). If $\delta(t_p) = \omega$, we put $c_p = j - i$. Note that we can safely assume that $c_p \leq |Q|$.

For every $p(i) \in Q_\square \times \mathbb{N}$, the strategy $\hat{\sigma}$ selects the configuration $q(j)$ obtained by applying the transition $t_p$ to $p(i)$. One can easily check that $\hat{\sigma}$ is $\mathcal{S}(Z)$-winning in every configuration of $Win(\square, \mathcal{S}(Z))$.

The construction of $\hat{\pi}$ is slightly more complicated, because player $\diamond$ must also make some progress in reaching a configuration of $Z$. Let $p \in Q_\diamond$ and let $i \in \mathbb{N} \cup \{\omega\}$ be the maximal index such that $p(i) \in Win(\diamond, \mathcal{R}(Z))$. We show that there is a transition $t_p \in T$ such that $\alpha(t_p) = p$ and player $\diamond$ still has an $\mathcal{R}(Z)$-winning strategy in $p(i)$ after deleting all outgoing transitions of $p$ except for $t_p$. Note that the existence of $\hat{\pi}$ easily follows from this claim, because then we can successively construct such a transition for every control state of $Q_\diamond$ (in polynomial time), and thus obtain the desired strategy $\hat{\pi}$.

To prove the claim, it suffices to consider the case when $i \neq \omega$ (if $i = \omega$, we apply Lemma 19). Realize that there must be some strategy $\pi$ which is $\mathcal{R}(Z)$-winning for player $\diamond$ in $p(i)$ and for every strategy $\sigma$ of player $\square$ we have that the resulting play does not visit a configuration $p(j)$ where $j \geq i$ (if there was no such $\pi$, player $\square$ could easily defeat every $\mathcal{R}(Z)$-winning strategy $\pi$, which is a contradiction). Let us fix such a $\pi$, and let $t_p$ be the transition which induces the edge $p(i) \mapsto \sigma(p(i))$. Further, for each $j < i$, let $\pi^j$ be a strategy defined by $\pi^j(q(k)) = \pi(q(k+i-j))$. We show that player $\diamond$ still has an $\mathcal{R}(Z)$-winning strategy $\bar{\pi}$ for $p(i)$ when all outgoing transitions of $p$ except for $t_p$ are deleted. Consider the strategy $\bar{\pi}$ obtained by applying the following rule recursively: "If a configuration of the form $p(j)$ is visited, the strategy $\bar{\pi}$ behaves like $\pi^j$ until another configuration of the form $p(m)$ is visited or a configuration with zero counter is reached." Note that $\bar{\pi}$ is $\mathcal{R}(Z)$-winning in $p(i)$, because a configuration of the form $p(j)$ can be revisited at most $i$ times in every play initiated in $p(i)$. □

As an immediate collary to Lemma 21 and Lemma 22, we obtain the following:

**Theorem 14.** *The problem whether $p(i) \in Win(\diamond, \mathcal{R}(Z))$ is in **P**. Further, there are counterless strategies $\sigma$ and $\pi$ such that $\sigma$ is winning for player $\square$ in every configuration of $Win(\square, \mathcal{S}(Z))$ and $\pi$ is winning for player $\diamond$ in every configuration of $Win(\diamond, \mathcal{R}(Z))$. The tuple of all $Win_\diamond(Q, i)$, $\sigma(p)$, and $\pi(q)$, where $0 \leq i \leq m_C$, $p \in Q_\square$, and $q \in Q_\diamond$, is constructible in time polynomial in $|\mathcal{M}|$.*

Now we turn our attention to $Z_C$ objectives and prove Theorem 13.

**Lemma 23.** *The problem whether $p(i) \in Win(\diamond, \mathcal{R}(Z_C))$ is **PSPACE**-complete.*

*Proof.* The **PSPACE** lower bound is obtained by reducing the emptiness problem for alternating finite automata (AFA) with one letter alphabet, which is known to be **PSPACE** complete [16] (see also [21] for a simpler proof). Intuitively, player $\diamond$ first increases the counter sufficiently and thus selects the word which should by accepted by a given AFA $\mathcal{A}$. The computation of $\mathcal{A}$ on the chosen word is then simulated by both players (the states of $\mathcal{A}$ are encoded in the finite control of the constructed VASS), and the counter is decreased after simulating one computational step. Player $\diamond$ aims to show that the chosen word is accepted by $\mathcal{A}$, which means that he wants to reach zero level in one of the control states that correspond to the accepting states of $\mathcal{A}$. Hence, the language accepted by $\mathcal{A}$ is nonempty iff player $\diamond$ has an $\mathcal{R}(Z_C)$-winning strategy in a configuration $p(1)$, where the set $C$ encodes the set of accepting states of $\mathcal{A}$.

The **PSPACE** upper bound follows also easily by employing the result of [31] which says that the emptiness problem for alternating two-way parity word automata (2PWA) is in **PSPACE**. A given eVASS game $G_\mathcal{M}$ with $\mathcal{R}(Z_C)$ objectives initiated in $p(i)$ can be easily simulated by a 2PWA $\mathcal{A}$ which tries to accept the infinite word $01^\omega$. Intuitively, the automaton $\mathcal{A}$ first performs $i$ steps to the right to simulate the initial counter value. The finite control of $\mathcal{M}$ is encoded in the states of $\mathcal{A}$ (the control states of $\mathcal{A}$ corresponding to $Q_\diamond$ are existential, and the control states corresponding to $Q_\square$ are universal; all of these control states have a non-accepting parity). The increment/decrement of the counter value is simulated by going right/left. If $\mathcal{A}$ reads 0, it enters an infinite loop in a special control state whose parity is accepting or non-accepting, depending on whether the corresponding control state of $\mathcal{M}$ belongs to $C$ or not, respectively. The $\omega$-transitions are implemented by allowing the automaton to go arbitrarily far to the right in a special control state, which is either existential or universal and has non-accepting or accepting parity, depending on whether the corresponding $\omega$-transition is performed by player $\diamond$ or player $\square$, respectively. At any moment, the automaton can switch back to the mode when it simulates the execution of $G_\mathcal{M}$. It follows that the only way how $\mathcal{A}$ can accept the word $01^\omega$ is to enter 0 in a "good" state which corresponds to a control state of $C$. Hence, player $\diamond$ has an $\mathcal{R}(Z_C)$ winning strategy in $p(i)$ iff $\mathcal{A}$ accepts the word $01^\omega$. $\square$

According to Lemma 23, the numbers $m_C, n_C$ and the tuple of all $\mathit{Win}_\diamond(C, i)$, where $0 \leq i < m_C+n_C$, are constructible in exponential time. Now we show that winning strategies for both players are finitely representable.

**Lemma 24.** *There is a strategy $\sigma$ for player $\square$ which is winning in every configuration of $\mathit{Win}(\square, \mathcal{S}(Z_C))$, and for all $p \in Q_\square$ and $i \geq m_C$ we have that*

$$\sigma(p(i)) \quad = \quad \sigma(p(m_C + ((i - m_C) \mod n_C)))$$

*Moreover, the value of all $\sigma(p(i))$, where $p \in Q_\square$ and $0 \leq i < m_C+n_C$, is computable in exponential time.*

*Proof.* Again, it suffices to ensure that $\sigma$ never leaves the winning region of player $\square$. Due to the ultimate periodicity of $\mathit{Win}(\square, \mathcal{S}(Z_C))$, the strategy $\sigma$ can be chosen so that for all $p \in Q_\square$ and $i \geq m_C$ we have that $\sigma(p(i)) = \sigma(p(m_C + ((i - m_C) \mod n_C)))$. Obviously, the value of $\sigma(p(i))$, where $p \in Q_\square$ and $0 \leq i < m_C+n_C$, is computable in exponential time because the sets $\mathit{Win}_\diamond(C, i)$, where $0 \leq i < m_C+n_C$, are computable in exponential time.
$\square$